\documentclass[11pt,a4paper]{article}
\usepackage{jcappub}
\usepackage{amsmath,amssymb}
\usepackage{natbib}
\usepackage{aas_macros}
\usepackage{verbatim}

\def\lbldef#1#2{\expandafter\gdef\csname #1\endcsname {#2}}

\def\href#1#2{#2}

\usepackage{graphicx}
\usepackage{epsfig}
\usepackage{amsmath}
\usepackage{amssymb}

\title{A 6\% measurement of the Hubble parameter at $z\sim0.45$: direct evidence of the epoch of cosmic re-acceleration}

\author[1,2]{Michele Moresco,}
\author[2]{Lucia Pozzetti,}
\author[1]{Andrea Cimatti,}
\author[3,4,5,6]{Raul Jimenez,}
\author[7]{Claudia Maraston,}
\author[3,4,5,9]{Licia Verde,}
\author[7]{Daniel Thomas,}
\author[1,2]{Annalisa Citro,}
\author[8]{Rita Tojeiro,}
\author[7]{and David Wilkinson}

\affiliation[1]{Dipartimento di Fisica e Astronomia, Universit\`a di Bologna, V.le Berti Pichat, 6/2, 40127, Bologna, Italy}
\affiliation[2]{INAF - Osservatorio Astronomico di Bologna, via Ranzani 1, 40127 Bologna, Italy}
\affiliation[3]{ICREA \& ICC, University of Barcelona (IEEC-UB), Barcelona 08028, Spain}
\affiliation[4]{Radcliffe Institute for Advanced Study, Harvard University, MA 02138, USA}
\affiliation[5]{Institute for Theory \& Computation, Harvard University, 60 Garden Street, Cambridge, MA 02138}
\affiliation[6]{Institute for Applied Computational Science, Harvard University, MA 02138, USA}
\affiliation[7]{Institute of Cosmology and Gravitation, Dennis Sciama Building, University of Portsmouth, Burnaby Road, Portsmouth, PO1 3FX, UK}
\affiliation[8]{School of Physics \& Astronomy, University of St Andrews, KY16 9SS, UK}
\affiliation[9]{Institute of Theoretical Astrophysics, University of Oslo, 0315 Oslo, Norway}
 
 \emailAdd{michele.moresco@unibo.it}
 \emailAdd{lucia.pozzetti@oabo.inaf.it}
 \emailAdd{a.cimatti@unibo.it}
 \emailAdd{rauljimenez@g.harvard.edu}
 \emailAdd{claudia.maraston@port.ac.uk}
 \emailAdd{liciaverde@icc.ub.edu}
 \emailAdd{daniel.thomas@port.ac.uk}
 \emailAdd{annalisa.citro@unibo.it}
 \emailAdd{rmftr@st-andrews.ac.uk}
 \emailAdd{david.wilkinson@port.ac.uk}

\abstract{Deriving the expansion history of the Universe is a major goal of modern cosmology. To date, the most accurate measurements have been obtained
with Type Ia Supernovae (SNe) and Baryon Acoustic Oscillations (BAO), providing evidence for the existence of a transition 
epoch at which the expansion rate changes from decelerated to accelerated. However, these results have been obtained 
within the framework of specific cosmological models that must be implicitly or explicitly assumed in the measurement.
It is therefore crucial to obtain measurements of the accelerated expansion of the Universe independently of
assumptions on cosmological models. Here we exploit the unprecedented statistics provided by the Baryon Oscillation 
Spectroscopic Survey (BOSS, \cite{Schlegel2009,Eisenstein2011,Dawson2013}) Data Release 9 to provide new constraints 
on the Hubble parameter $H(z)$ using the {\em cosmic chronometers} approach. We extract a sample of more than 130000 of the most massive 
and passively evolving galaxies, obtaining
five new cosmology-independent $H(z)$ measurements in the redshift range $0.3<z<0.5$, with an accuracy of $\sim$11--16\% 
incorporating both statistical and systematic errors. Once combined, these measurements yield a 6\% accuracy constraint of 
$H(z=0.4293)=91.8\pm5.3$ km/s/Mpc.
The new data are crucial to provide the first cosmology-independent determination of the transition redshift at high statistical significance, 
measuring $\rm z_{t}=0.4\pm0.1$, and to significantly disfavor the null hypothesis of no transition
between decelerated and accelerated expansion at 99.9\% confidence level.
This analysis highlights the wide potential of the cosmic chronometers approach: it permits to derive constraints on the expansion 
history of the Universe with results competitive with standard probes, and most importantly, being the estimates independent of 
the cosmological model, it can constrain cosmologies beyond -- and including -- the $\Lambda$CDM model.}

\begin{document}

\maketitle
\flushbottom

\section{Introduction}
Almost 15 years have now passed since the discovery of the accelerated expansion of the Universe, based on the
work by refs. \cite{Riess1998} and \cite{Perlmutter1999}. Since then, constraining the expansion rate of the Universe as a function
of redshift $H(z)$ has become one of the most compelling tasks of modern cosmology, since it determines the scale factor $a(t)$ in 
the Friedmann-Lemaitre-Robertson-Walker (FLRW) metric; as a consequence, this allows us to probe the properties of the fundamental 
components of the Universe, helping to better understand their nature. 
Widening the panorama of cosmological probes is therefore of extreme interest, in order to take advantage of the strength of 
each method and to keep systematics under control (for a detailed review, see ref.~\cite{Weinberg2013}).

So far, the best measurements have been obtained with standard candles (SNe) \cite{Riess1998,Perlmutter1999} 
and standard rulers (BAO) \cite{Eisenstein2005,Cole2005}. These cosmological probes have provided exceptional results over the last 
decades, and contributed, together with the study of the Cosmic Microwave 
Background (CMB) \cite{Planck2015}, to the development of the standard $\Lambda$CDM cosmological model. 
One of their main limitations, however, is that they do not constrain the Hubble
parameter {\color{black}directly}, but one of its integrals (e.g., the luminosity distance \cite{Riess1998}) or the parameters that are used to model it.
In order to set constraints on the cosmological model, it is necessary to obtain an independent determination of the expansion 
rate, which could then be used to test the model itself.

A possible way to achieve this task is given by the ``cosmic chronometers'' method. According to this method, firstly suggested by 
ref.~\cite{Jimenez2002}, the relative age of old and passive galaxies $dz/dt$ can be used to directly constrain the expansion history of the Universe. The 
most important point to emphasize is that because the differential dating of passively evolving galaxies only depends on atomic 
physics and does not include any integrated distance measurement over redshift, it is independent of the cosmological 
model or assumptions about the metric, and thus can be used to place constraints on it.

In ref.~\cite{Moresco2012a}, analyzing a sample of $\sim$11000 massive and passive galaxies, eight new measurements of the 
Hubble parameter have been provided with an accuracy of $\sim$5-12\% in the redshift range $0.15<z<1.1$, significantly extending the redshift 
coverage and precision of previous similar analysis \cite{Simon2005,Stern2010}. The majority of the sample in this work was at low redshift 
($z<0.3$), where the most accurate constraints were obtained. The potential of this new method in comparison with more standard probes has been 
studied by many authors \cite{Moresco2012b, Zhao2012,Wang2012, Sorensen2013}, which demonstrated how the cosmic chronometers method 
can be competitive for many aspects with Supernovae type Ia (SNe) and Baryon Acoustic Oscillation (BAO) in constraining cosmological parameters.

In this paper, we want to exploit the phenomenal set of data provided by the SDSS-III Baryon Oscillation Spectroscopic Survey (BOSS, 
\cite{Schlegel2009,Eisenstein2011,Dawson2013}), which represents the largest sample of massive galaxies spectra available so far at $0.2<z<0.8$, to provide 
new $H(z)$ measurements with the highest possible accuracy at these redshifts. This sample allows us to minimize all possible 
uncertainties, by improving the statistics used so far by $\sim$2 orders of magnitude at these redshifts. 
BOSS data also represent a very well defined spectroscopic sample, being targeted to select the most massive envelope of the galaxy population at these 
redshifts \cite{Maraston2013}. Moreover, the redshift range probed by BOSS proves to be fundamental to set cosmology independent 
constraints on the redshift at which the Universe expansion turns from decelerated to accelerated, which we refer to as the cosmological 
transition redshift \cite{Riess2007,Lima2012,Capozziello2014,Capozziello2015}.

This paper is organized as follows. In section \ref{sec:method} we introduce the technique, its potential drawbacks and how they are addressed. In
section \ref{sec:data} we describe how the dataset has been selected and describe its properties. In section \ref{sec:analysis} we
provide the stellar metallicity constraints for our sample. We also discuss how the relation between parameters and observables used
in the method has been obtained and calibrated. Finally, in section \ref {sec:results} we present our results, providing five new $H(z)$
points in the redshift range $0.35<z<0.5$, and using them, in combination with available literature data, to obtain new constraints
on the cosmological transition redshift.

\section{Method: the cosmic chronometers approach}
\label{sec:method}
An accurate measurement of the expansion rate as a function of cosmic time is extremely challenging. 
An alternative and promising method is the ``cosmic chronometers'' approach, which uses the fact that 
the expansion rate can be expressed as $H(z) ={\dot a}/a = -1/(1+z)\, dz/dt$.
Since the quantity $dz$ is obtained from spectroscopic surveys with high accuracy, the only quantity to be measured is the differential 
age evolution of the Universe ($dt$) in a given redshift interval ($dz$). Therefore, a measurement of $dt$ corresponds to a direct 
cosmology-independent measurement of the Hubble parameter. 
There are two main challenges to be faced: the identification of an optimal tracer of the aging of the Universe 
with redshift (a ``cosmic chronometer''), and the reliable dating of its age.

The best cosmic chronometers are galaxies that are evolving passively on a timescale much longer than their age difference. 
Based on several observational results, there is general agreement that these are massive (${\cal M}_{\rm stars}>10^{11}
{\cal M}_{\odot}$) early-type galaxies which formed the vast majority ($>$90\%) of their stellar mass {\color{black} rapidly ($\sim$0.1-0.3 Gyr) early in the Universe
(at high-redshifts $z >2-3$), and have not experienced any subsequent major episode of star formation since. Therefore they are
the oldest objects at any redshift \cite{Treu2005,Renzini2006,Thomas2005,Thomas2010,Pozzetti2010,Conroy2014,
McDermid2015}. Thus, when observed at cosmic times considerably later than their formation epoch, the age evolution of their stars serves a clock 
that is synchronized with the evolution of cosmic time. 
Previous works \cite{Bender1998,Carson2010,Moresco2011,Moresco2012a} have also demonstrated that it is possible to obtain reliable and accurate cosmic 
chronometers using passively evolving selected galaxies. Recently, the independent analysis of SDSS-DR8 luminous red galaxies by ref.~\cite{Liu2015} has 
confirmed that massive and quiescent galaxies can be reliably used as cosmic chronometers.
  
It is important to underline that the main strength of this method is that it relies on a {\em differential approach}. On the one hand, it should be noted that the relevant 
quantities in this approach are the relative ages $dt$, which have the advantage of factorizing out systematic effects inherent to absolute age estimates. 
On the other hand, this method allows us to keep under control many observational biases which may 
affects this analysis. One of the main issues is the so-called progenitor bias \cite{Franx1996,Vandokkum2000}, for which 
high-redshift ETGs are biased by sampling only the oldest and more massive progenitors of more local galaxies, 
therefore changing the slope of the age-redshift relation. This effect is more severe when galaxy samples are 
compared in wide redshift ranges, but here each 
$H(z)$ value is estimated within small redshift slices, with $\Delta z=0.1$ (as discussed in section \ref{sec:Hz})
corresponding to a difference in cosmic time of approximatively 0.7 Gyr at $z\sim0.45$. This is a rather short time for a 
potential significant evolution of these massive and passive systems, in particular considering that 
on average they would require more than the age of the Universe to double their mass \cite{Pozzetti2010,Moresco2013}. 
Moreover, the progenitor bias dominates the low-mass range of the distribution, whereas BOSS galaxies have been selected 
to be the most massive galaxies ($\rm log(M/M_{\odot})\gtrsim11$) between $0.3<z<0.7$ (see Fig. 1 of 
ref.~\cite{Maraston2013}), and further divided in bin of velocity dispersion, which is a proxy for stellar mass.

To minimize the dependence of the age estimate on evolutionary stellar population synthesis (EPS) models, refs.~\cite{Moresco2011,Moresco2012a} proposed an improvement 
to this technique, consisting in studying a direct observable of galaxy spectra, the 4000~\AA~ break ($D4000$), instead of galaxy ages. 
The $D4000$ feature is a break in the observed spectrum of galaxies defined as the ratio between the continuum flux densities $\langle F_{\nu}\rangle$ in a red band 
and a blue band around 4000~\AA~restframe \cite{Bruzual1983}. The break originates from the onset of a series of metal absorption features, and is known to correlate 
with the stellar metallicity and age of the stellar population \cite{Hamilton1985}, and to be less dependent on the star formation history (SFH) for old stellar populations. 
Different choices have been proposed to measure the $D4000$, changing 
the range of the red and blue bands \cite{Bruzual1983,Hamilton1985,Balogh1999}. In this work, we considered the definition with narrower bands ($D_{n}4000$, 
3850--3950~\AA~and 4000--4100~\AA), since it has been shown to be less affected by potential reddening effects \cite{Balogh1999}.
The assumption of a linear relation between $D4000$ and age of a galaxy has been proven to be an extremely good approximation for various $D4000$ ranges 
(see refs.~\cite{Moresco2011,Moresco2012a}), i.e.
\begin{equation}
D_{n}4000=A(SFH, Z/Z_{\odot})\cdot {\rm age} + B ,
\label{eq:D4000age}
\end{equation}
where $A(SFH, Z/Z_{\odot})$ is the slope of the $D_{n}4000$--age relation, and $B$ its normalization. From eq.~\ref{eq:D4000age}, it is therefore possible to express the 
Hubble parameter as:
\begin{equation}
H(z)=-\frac{1}{1+z} A(SFH, Z/Z_{\odot}) \frac{dz}{dD_{n}4000} ,
\label{eq:HzD4000}
\end{equation}
where now statistical and systematic effects have been decoupled. The factor $dz/dD_{n}4000$ is only dependent 
on observables, while degeneracies between parameters or assumptions of EPS models 
are contained in the factor $A(SFH, Z/Z_{\odot})$. The main source of systematic errors is the adopted EPS model that quantifies the dependence of the  
$D_{n}4000$ on age, metallicity Z and SFH. In Sect. \ref{sec:D4000meas} we will discuss the robustness of the assumption in Eq. \ref{eq:D4000age},
as well as the detailed dependence of the factor $A(SFH, Z/Z_{\odot})$ on metallicity and SFH.

In ref.~\cite{Moresco2012a} it was demonstrated that this new approach, which relies on the spectroscopic differential evolution of cosmic chronometers, is extremely 
robust against the choice of EPS models, and does not strongly depend on the SFH assumptions, the reason being that the selection criteria adopted significantly reduce the 
possibility of having prolonged SFHs, and spectra are usually well-fit with models with short star formation bursts.

\section{Data and sample selection}
\label{sec:data}

The BOSS survey was designed to accurately map the clustering of large-scale structures up to $z\sim0.7$, and in particular to measure the baryonic acoustic oscillation scale 
and constrain the expansion history of the Universe. It collected both photometric and spectroscopic
data for $\sim$1.5 million galaxies over approximatively 10000 square degrees. The sample comprises the standard $ugriz$ SDSS photometry;
the spectroscopic sample has been photometrically targeted with a color cut in the $(g-r)$-$(r-i)$ plane to select the most massive galaxies 
within a passively evolving population, and with a nearly constant number density as a function of redshift up to $z\sim0.7$ \cite{White2011,Eisenstein2011}.
BOSS spectra are obtained at a resolution $\rm R\sim2000$ in the observed wavelength range 3750--10000~\AA. Stellar masses for a large BOSS sample 
have been obtained by ref. \cite{Maraston2013} through a best fit of the observed photometry. Throughout this work, we considered the estimates obtained 
assuming a Kroupa initial mass function, accounting for stellar mass losses from stellar evolution. Ref. \cite{Thomas2013} estimated 
emission lines properties and stellar velocity dispersions on the same sample.

Starting from BOSS Data Release 9 (DR9), we considered all galaxies with measured stellar masses, emission lines and stellar velocity dispersions, 
obtaining an original sample consisting of 848697 galaxies for which the $D_{n}4000$ index has been measured. 

\begin{figure}[t!]
\begin{center}
\includegraphics[angle=0, width=0.45\textwidth]{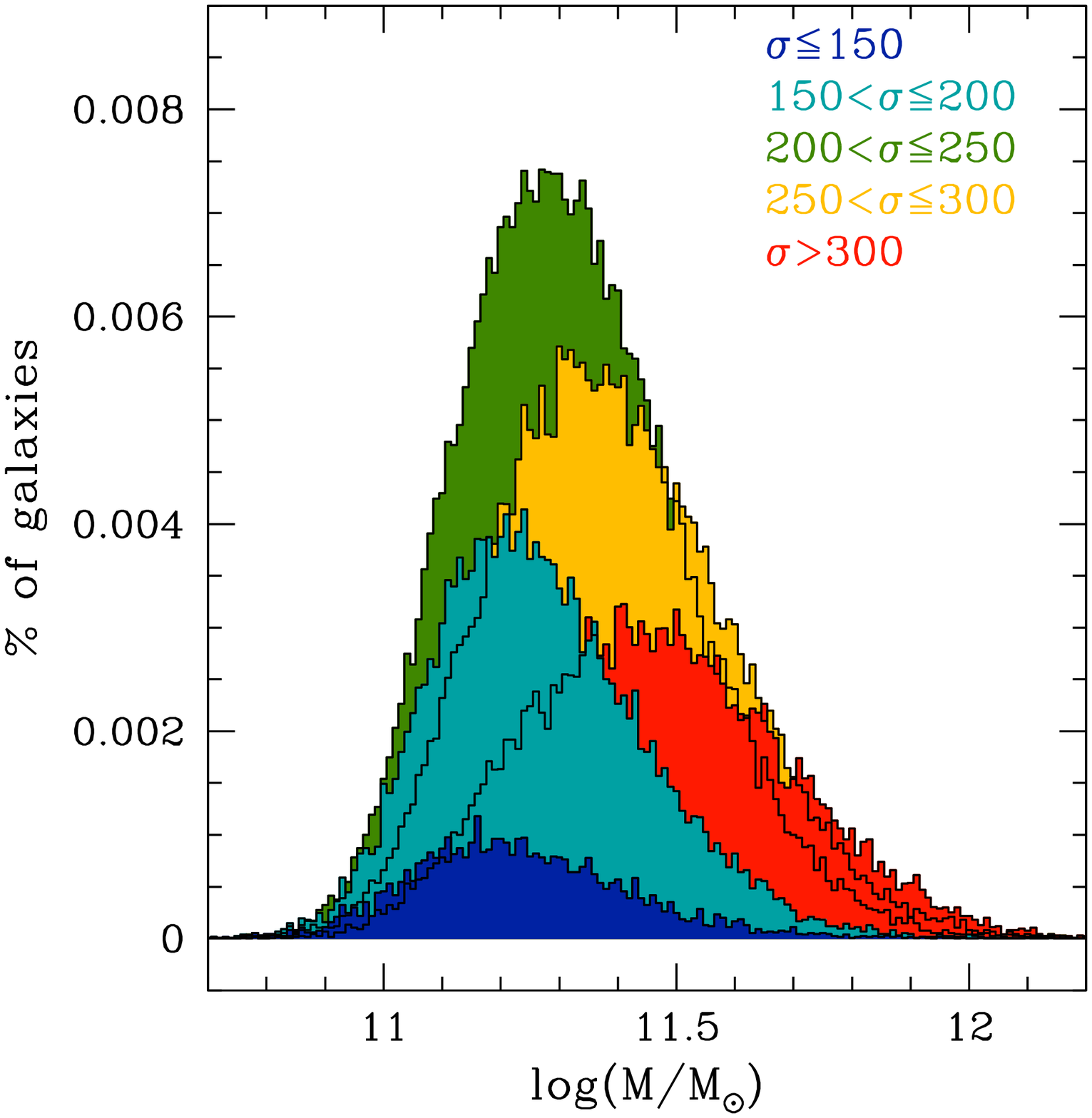}
\includegraphics[angle=0, width=0.45\textwidth]{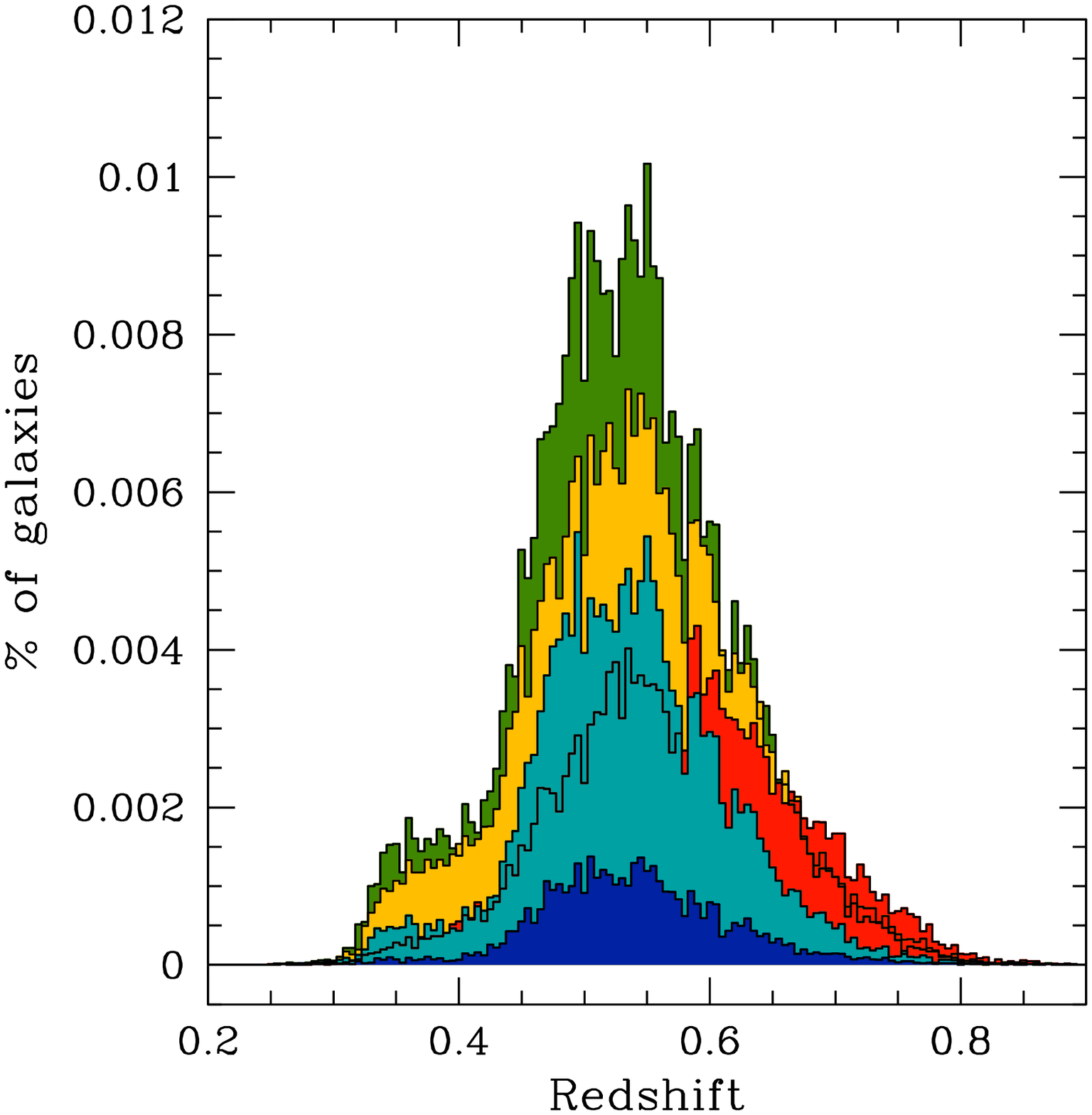}
\caption{Stellar mass (left panel) and redshift (right panel) distributions, colored by the various velocity dispersion subsamples.
\label{fig:masshist}}
\end{center}
\end{figure}

As discussed in section \ref{sec:method}, the ``cosmic chronometers'' method relies on the selection of an optimal sample of massive and passive galaxies, minimally 
contaminated by star-forming galaxies outliers. A star-forming population, with its evolving properties (in terms of age, mass, metal content) as a function of cosmic time, 
could potentially bias the results. On the contrary, as already discussed, the most massive and passive galaxies represent the most homogeneous sample in terms of
epoch and timescale of formation, thus being the best tracers of the age of the Universe as a function of redshift. It is therefore of utmost importance to ensure the purity of the
sample by removing all galaxies with residual on-going star formation. A combination of selection cuts in color, spectroscopic properties, and stellar mass 
(${\cal M}_{\rm stars} \gtrsim 10^{10.75} {\cal M}_{\odot}$) has been demonstrated to maximize the purity of such a passively evolving galaxy sample \cite{Moresco2013}.\\
\begin{table}[t!]
\begin{center}
\begin{tabular}{cccccc}
\hline
\hline
subsample & redshift &$\sigma$ &$\rm log(M/M_{\odot})$ & S/N($D_{n}4000$) & \# of\\
 $\rm[km/s]$ & range &[km/s] & & & galaxies\\
\hline
$\sigma\le150$ & $0.45-0.55$ & 133 $\pm$ 17 & 11.22 $\pm$ 0.18 & $11.8^{+51.2}_{-5.1}$ & 5557 \\
$150<\sigma\le200$ & $0.32-0.65$ & 182 $\pm$ 16 & 11.25 $\pm$ 0.18 & $11.9^{+53.9}_{-4.9}$ & 22710 \\
$200<\sigma\le250$ & $0.33-0.75$ & 226 $\pm$ 18 & 11.31 $\pm$ 0.19 & $12.1^{+53.9}_{-4.9}$ & 45995 \\
$250<\sigma\le300$ & $0.35-0.8$ & 271 $\pm$ 17 & 11.37 $\pm$ 0.21 & $12.4^{+54.2}_{-5}$ & 36958 \\
$\sigma>300$ & $0.45-0.85$ & 331 $\pm$ 32 & 11.46 $\pm$ 0.22 & $12.6^{+57.8}_{-5.2}$ & 22677 \\
\hline
\hline
\end{tabular}
\caption{Median properties of the sample measured on individual galaxies in each velocity dispersion subsamples.}
\label{tab:sample_prop}
\end{center}
\end{table}
\begin{figure}
\begin{center}
\includegraphics[angle=-90, width=0.98\textwidth]{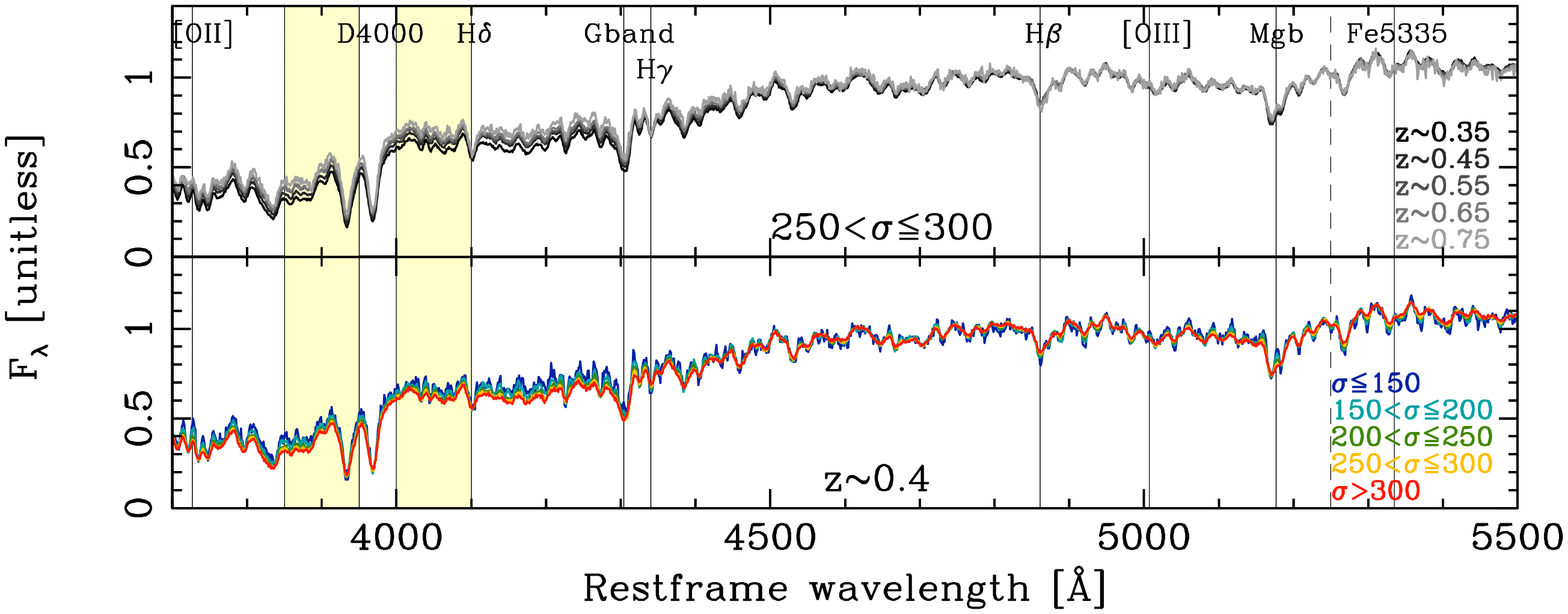}
\includegraphics[angle=-90, width=0.98\textwidth]{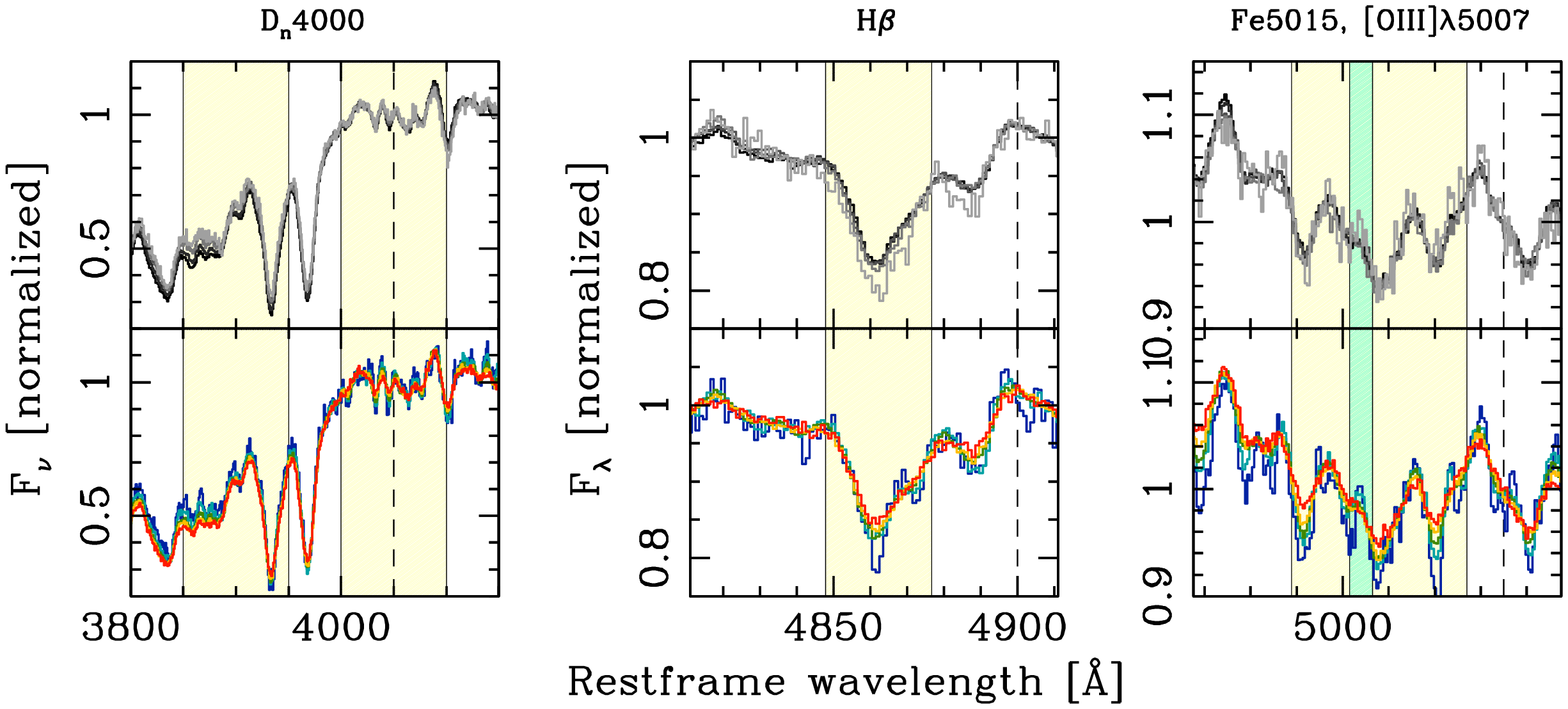}
\caption{Median stacked spectra as a function of redshift bins and $\rm\sigma$ bins (the upper and lower panels of each plot respectively). The spectra in the upper panels of
each plot are extracted at fixed $\rm\sigma$ (grey lines, ${\rm 250<\sigma<300}$), while the spectra in the lower panels are extracted at fixed redshift (colored lines, $z\sim0.4$). 
The upper plots show the median spectra. All spectra have been normalized near to the vertical dashed lines; therefore the differences in the upper panels 
flatten at higher wavelengths, but the steepening of the slope of the continuum with increasing mass and decreasing redshift is evident. This trend may be interpreted in the framework 
of the ``mass-downsizing'' scenario, with more massive galaxies being redder and older than less massive ones. 
The lower plots are zoom-in around three specific absorption features, i.e. $D_{n}4000$, H$\beta$, and Fe5015, highlighted in yellow; it is also shown, highlighted in green,
the region corresponding to [OIII]$\lambda$5007 line.
\label{fig:spectra}}
\end{center}
\end{figure}
We therefore applied the following cuts to obtain our sample.
\begin{itemize}
\item {\bf Color cut.} We selected galaxies with a color $\rm(g-i)>2.35$ \cite{Maraston2013}, which was proven to select `early-type' (passive) galaxies 
\cite{Masters2011}.
\item {\bf Emission line cut.} We selected galaxies without detectable emission lines, where the main cut has been done on the $\rm[OII]\lambda$3727 line, since, given the
wavelength range, it is measurable throughout the entire redshift range. We selected galaxies with an equivalent width (EW) $\rm EW([OII])<5$~\AA~and a 
signal-to-noise ratio (S/N(EW)) $<$ 2. The sample has been further 
cleaned by discarding galaxies with significant emission in $\rm H\alpha$, $\rm[OIII]\lambda$5008 and $\rm H\beta$, namely at $S/N(EW)>2$.
\item {\bf $\mathbf{D_{n}4000}$ measurement cut.} To remove low-quality measurements corresponding to spectra with lower $S/N$, we selected galaxies with an error 
$\sigma(D_{n}4000)<0.25$, obtaining on average a S/N$(D_{n}4000)>10$.
\end{itemize}
The final sample contains 133912 galaxies, and represents the largest sample of massive and passive 
galaxies to date, improving current statistics by $\sim$2 orders of magnitude at these redshifts. 
This sample was further divided into 5 velocity dispersion bins: $\sigma<150$, 
$150<\sigma<200$, $200<\sigma<250$, $250<\sigma<300$, and $\sigma>300$ km/s. The properties of the various subsamples are reported in 
Tab. \ref{tab:sample_prop}, and the stellar mass and redshift distributions are shown in figure \ref{fig:masshist}.

Median stacked spectra have been created for each subsample by co-adding individual spectra in redshift bins of $\Delta z=0.025$, and the results are shown in figure \ref{fig:spectra}.
The redshift binning has been optimized by choosing the smallest width for which the median $D_{n}4000$--z were not noise dominated; the choice of a small binning allows to follow
the redshift evolution without excessive smoothing.

The stacked spectra presented in figure \ref{fig:spectra} are typical of a passive population, with a clear red continuum, characteristic absorption features and no noticeable emission lines, 
demonstrating the effectiveness of the implemented selection criteria.
A zoom-in around three specific lines is shown for $D_{n}4000$, H$\beta$, and Fe5015 respectively. The H$\beta$ absorption line is a feature which is almost independent of metallicity, 
depending mostly only on the age of the galaxy population, being shallower for high ages. From the visual inspection of this line the aging of galaxies with cosmic time is evident, with 
galaxies at higher redshifts being younger. Fe5015 is instead a good proxy of the metallicity of the galaxy; in this case no significant trend with redshift can be noticed, giving a hint of no metallicity 
evolution of this population with cosmic time. The absence of clearly visible [OIII]$\lambda$5007 line shows the purity of the sample, where all contamination from emission-line galaxies 
has been removed.

\section{Analysis}
\label{sec:analysis}
In this section we discuss the methods used to constrain the stellar metallicity Z and to measure and calibrate the $D_{n}4000$--z relations.
\subsection{Constraining the stellar metallicity}
\label{sec:metallicity}
The stellar metallicity Z of the various subsamples has been estimated from full spectral fitting of the median stacked spectra, in bins of redshift and velocity dispersion.
The measurements have been performed on the median stacked spectra to increase the $S/N$ from a typical value of about $\rm5/\AA$ for individual galaxies \cite{Thomas2013} to 
about $\rm300-600/\AA$ around $\rm\lambda_{rest}=5000$~\AA. Depending on the considered bin, the final spectra were obtained by stacking from $\sim1500$ up to $\sim5000$ 
galaxies (Tab.~\ref{tab:sample_prop}). Moreover, in ref. \cite{Citro2016} it was verified that the measurement of the metallicity on individual spectra is consistent with the one obtained on 
median stacked spectra. It is worth also noting that ref.~\cite{MaStro} found a general agreement between estimates obtained from full spectral fitting and from other techniques 
(e.g. Lick indices), and these results have been more recently confirmed by ref.~\cite{Conroy2014}.

Different codes have been used to estimate the mass-weighted metallicity, and the results compared to quantify the robustness of the measurements, namely STARLIGHT \cite{Cid2005}, 
VESPA \cite{Tojeiro2007} and FIREFLY \cite{Wilkinson2015b}.
To further assess the dependence of the results on the assumed parameters, each code has been run adopting two different EPS model, Bruzual \& 
Charlot (2003) (hereafter BC03, \cite{BC03}) and Maraston \& Str{\"o}mb{\"a}ck (2011) (hereafter M11, \cite{MaStro}). These two models present substantial differences, 
such as the treatment of the thermally pulsating asymptotic giant branch phase, the method used to estimate integrated spectra, and the stellar evolutionary 
models adopted to build the isochrones (for more extensive discussion, see e.g. \cite{Maraston2006,MaStro}). Moreover, they are also based on independent libraries of stellar 
spectra, the latest MILES models \cite{Falcon2011} for M11 and STELIB \cite{LeBorgne2003} for BC03. The M11 models are based on a selection of libraries of empirical 
stellar spectra (Pickles, ELODIE, and STELIB), but we studied the ones with MILES models to be consistent with the work done in previous analysis \cite{Moresco2012a}.
The two models provide similar metallicity ranges for exploration: 
$Z/Z_{\odot}=[0.4,1,2.5]$ for BC03 and $Z/Z_{\odot}=[0.5,1,2]$ for M11, and have a similar resolution, 3~\AA~across the wavelength range from 3200~\AA~to 
9500~\AA~for BC03 \cite{BC03}, and 2.54~\AA~across the wavelength range from 3525~\AA~to 7500~\AA~for M11 \cite{Beifiori2011}, similar to BOSS 
spectra \cite{Smee2013}.\\ The full spectral fitting codes applied to estimate the metallicity are described here:
\begin{figure}[t!]
\begin{center}
\includegraphics[angle=-90, width=0.9\textwidth]{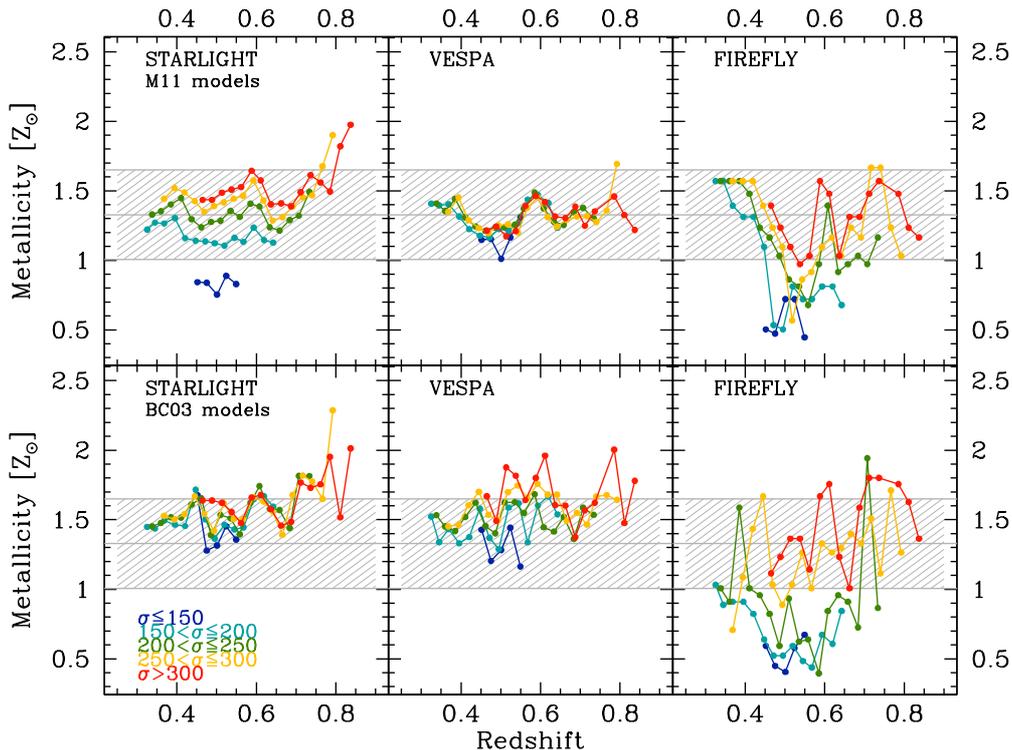}
\caption{Stellar metallicity estimated from full spectral fitting with STARLIGHT, VESPA and FIREFLY (respectively left, center and right panels),
and adopting M11 and BC03 models (respectively upper and lower panels). The grey shaded region represents the mean value, averaged
between all models and codes, as discussed in the text.
\label{fig:metFit}}
\end{center}
\end{figure}
\begin{itemize}
\item The full-spectrum fitting code STARLIGHT \cite{Cid2004,Cid2005} provides a fit to both the galaxy spectral continuum and spectral features. It simultaneously fits a stellar 
population mix, which is given as a combination of spectra defined in user-made libraries, the global stellar velocity dispersion and the amount of dust extinction (in terms of $A_{V}$). 
The contribution of each library spectrum to the best fit model is enclosed in the so-called light-fraction and mass-fraction population vectors, which contain, respectively, the light 
and the mass fractions with which each library model contributes to the best fit spectrum at a reference wavelength $\lambda$. The best fit model is then derived by exploring the
parameters space through a mixture of simulated annealing and Metropolis-Hasting Markov Chain Monte Carlo techniques.
\item VErsatile Spectral Analysis (VESPA, \cite{Tojeiro2007,Tojeiro2009}) is a full spectral fitting code to recover non-parametric star-formation and metallicity histories from optical 
spectra. VESPA works on a grid of ages logarithmically binned between 0.02 and 14 Gyr with a resolution adaptable to the quality of the data, increasing the resolution only where 
the data demand it. VESPA imposes no constraints on the amount of star formation or metallicity in each age bin. VESPA finds the best-fit solution by appropriately linearizing the 
problem and then performing a single matrix inversion to solve the problem. Errors and full covariance matrices are computed by perturbing and then fitting 
the original best-fit solution a number of times. Dust attenuation is modeled according to either a one or two-parameter mixed-slab dust model of ref. \cite{Charlot2000}. 
\item FIREFLY \cite{Wilkinson2015b} is a full spectral fitting code designed to recover galaxy properties and their errors as a function of input model ingredients, 
such as stellar library, whilst also mapping out stellar population property degeneracies from both the inherent degeneracies in galaxy spectra and errors in the data. It constructs 
linear combinations of single burst in order to build up complex star formation histories. It has been tested to work well down to signal-to-noise of 5 \cite{Wilkinson2015b}, and 
includes an innovative method for treating the effects of dust attenuation and flux calibration through use of Fourier filters. The spectral fitting is adapted to the specific
velocity dispersion of the data.
\end{itemize}

The mass-weighted stellar metallicity estimated for the various ETG subsamples is shown in figure \ref{fig:metFit}. The various fits show on average a slightly over-solar metallicity, 
with a value $Z/Z_{\odot}\sim1-1.5$. This result is both expected on a theoretical basis, and confirmed by other independent analysis. On a theoretical basis, for galaxies which have fully
exhausted their gas reservoir, having completed their mass assembly at high redshift and being passively evolving since then, a negligible evolution in stellar metallicity is 
expected. Moreover refs.~\cite{Panter2003,Gallazzi2005,Panter2007,Panter2008,Thomas2010} analyzing the Sloan Digital Sky Survey at redshifts $0<z<0.3$ similarly found a 
slightly over-solar metallicity for galaxies with masses $\rm log(M/M_{\odot})\sim11$, and these measurements are confirmed also for galaxies of the same mass at much higher redshifts 
($z\sim$1.5--2 \cite{Toft2012,Onodera2015}).

While no statistically significant difference is detectable between the best-fits obtained with the different codes, FIREFLY presents a slightly larger dispersion,
that could be due to the fact that it does not assume, as STARLIGHT, a fixed velocity dispersion. A slight difference is present between BC03 and M11 estimates, with BC03 values 
preferring a metallicity $Z/Z_{\odot}\sim1.5$ while M11 $Z/Z_{\odot}\sim1.25$; this difference is due to the intrinsic different input physics and methods in the models, as discussed e.g. in 
refs.~\cite{MaStro,Maraston2006}. It is important to notice that all best-fits find a negligible metallicity evolution, in agreement with what found from the visual inspection of the 
median spectra (see figure \ref{fig:spectra}), confirming that galaxies have been evolving little over the redshift range probed.

To be as conservative as possible, we therefore decided to average the six metallicity estimates obtained with all models and codes, so that the dispersion of the obtained 
distribution takes all the systematic uncertainties into account. In this way, we obtain a mean value for the stellar metallicity of $Z/Z_{\odot}=1.35\pm0.3$. We verified
that this estimate is also stable when calculating the median, or restricting the redshift range, being always compatible within the errorbars.

\subsection{Measuring and calibrating the median $D_{n}4000$--z relations}
\label{sec:D4000meas}

For each velocity dispersion subsample, we calculated the median $D_{n}4000$ from the individual measurements with a $\Delta z=0.025$ redshift binning. 
The associated error has been estimated using the median absolute deviation (MAD), defined as $\mathrm{MAD}=1.482\cdot \mathrm{median}(|D_{n}4000-\mathrm{median}
(D_{n}4000)|)$), divided by $\sqrt{N}$, i.e. $\sigma_{med}(D_{n}4000)=\mathrm{MAD}/\sqrt{N}$ \cite{Hoaglin1983}. We verified that the median value obtained
from the individual galaxies is consistent with the value estimated on the median stacked spectrum in each bin.

\begin{figure}[t!]
\begin{center}
\includegraphics[angle=0, width=0.49\textwidth]{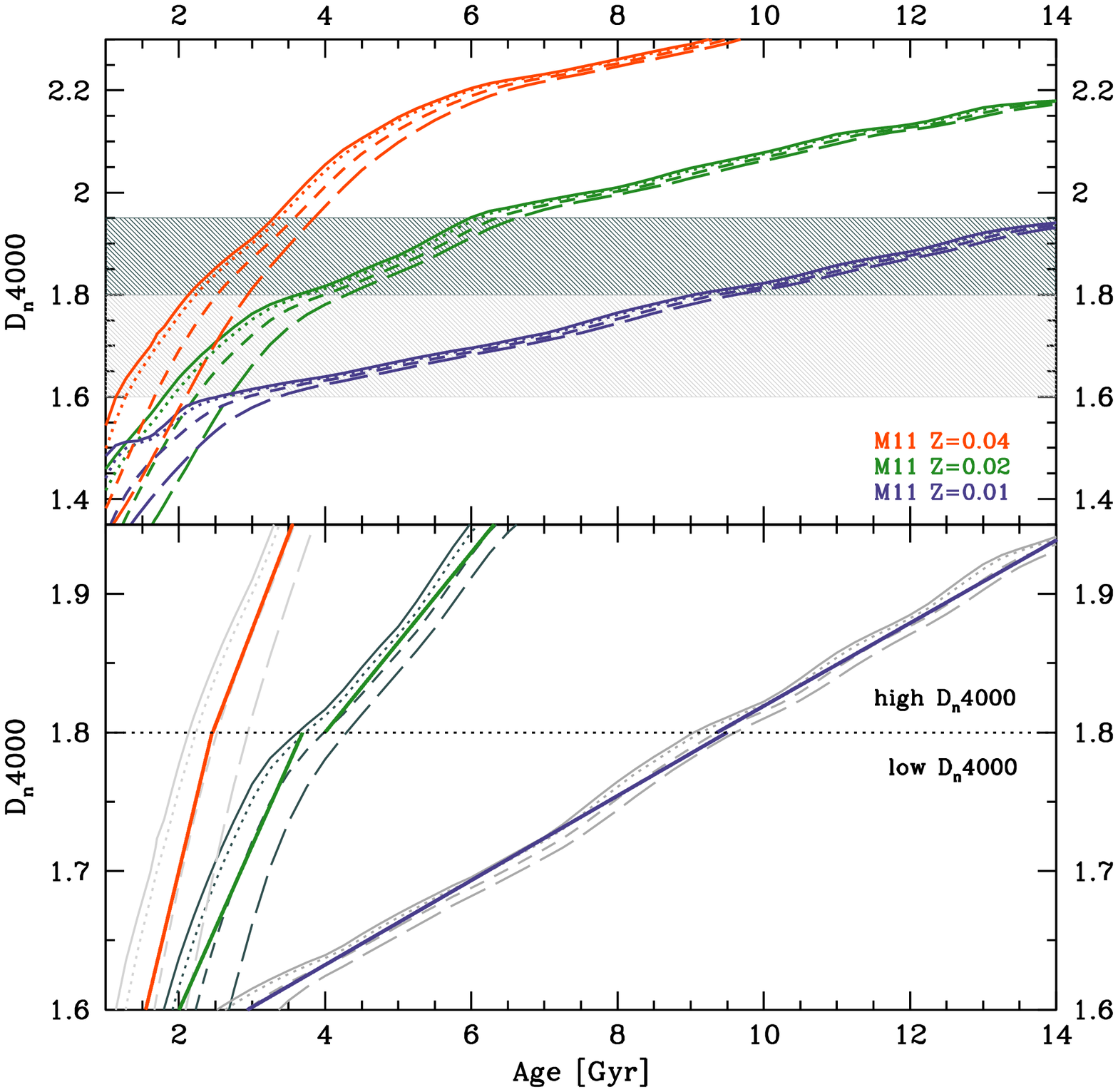}
\includegraphics[angle=0, width=0.49\textwidth]{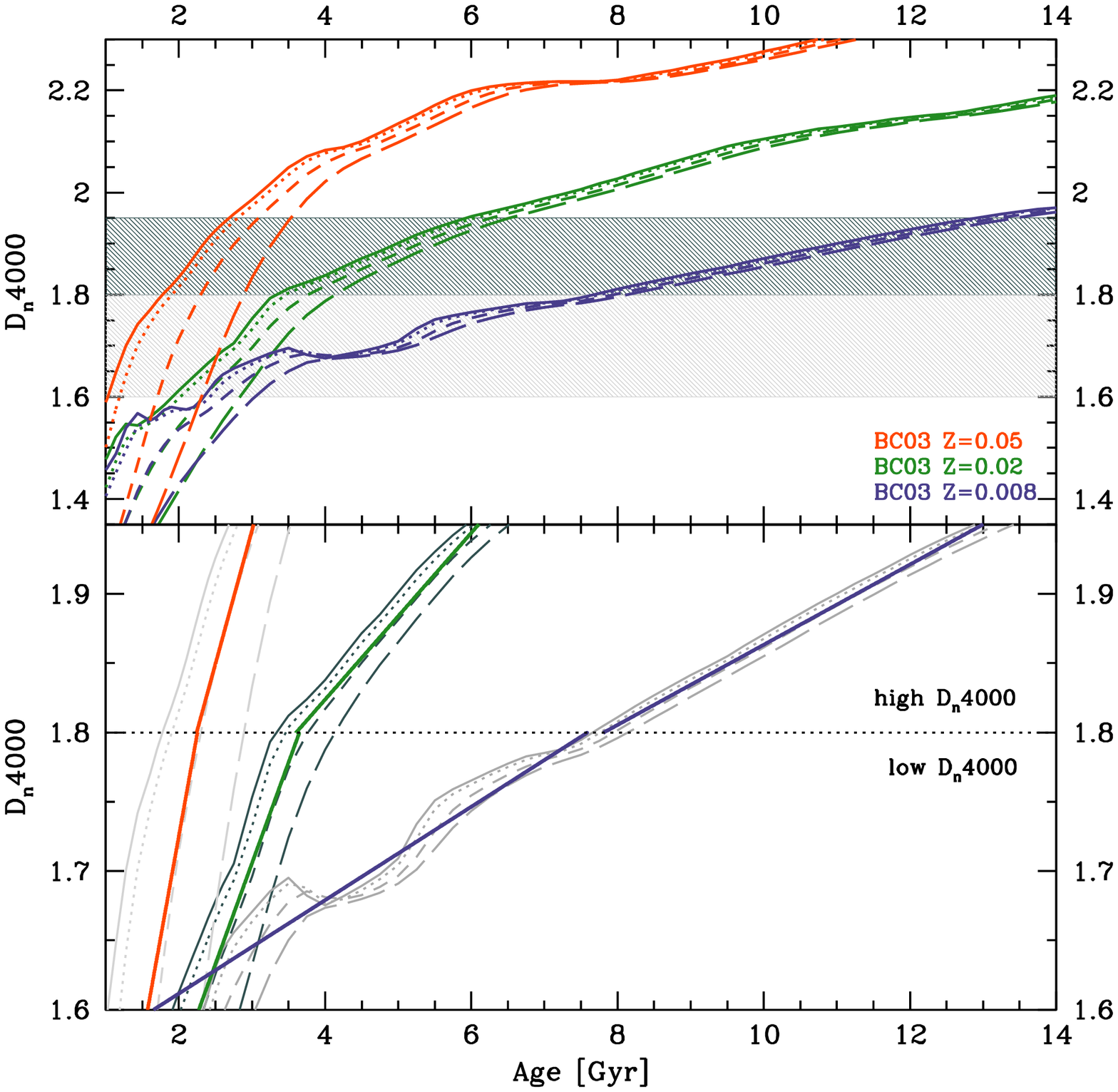}
\includegraphics[angle=0, width=0.49\textwidth]{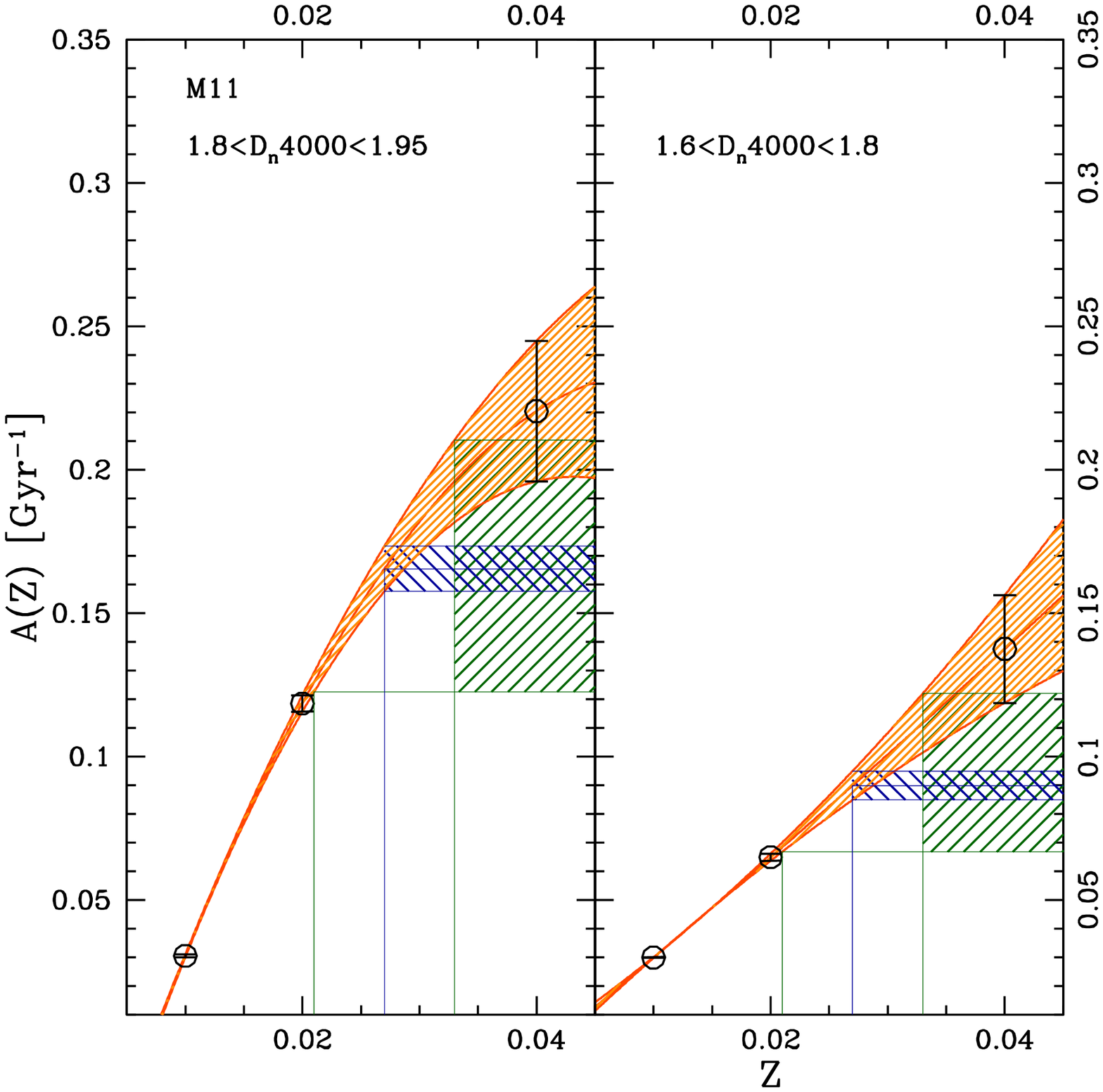}
\includegraphics[angle=0, width=0.49\textwidth]{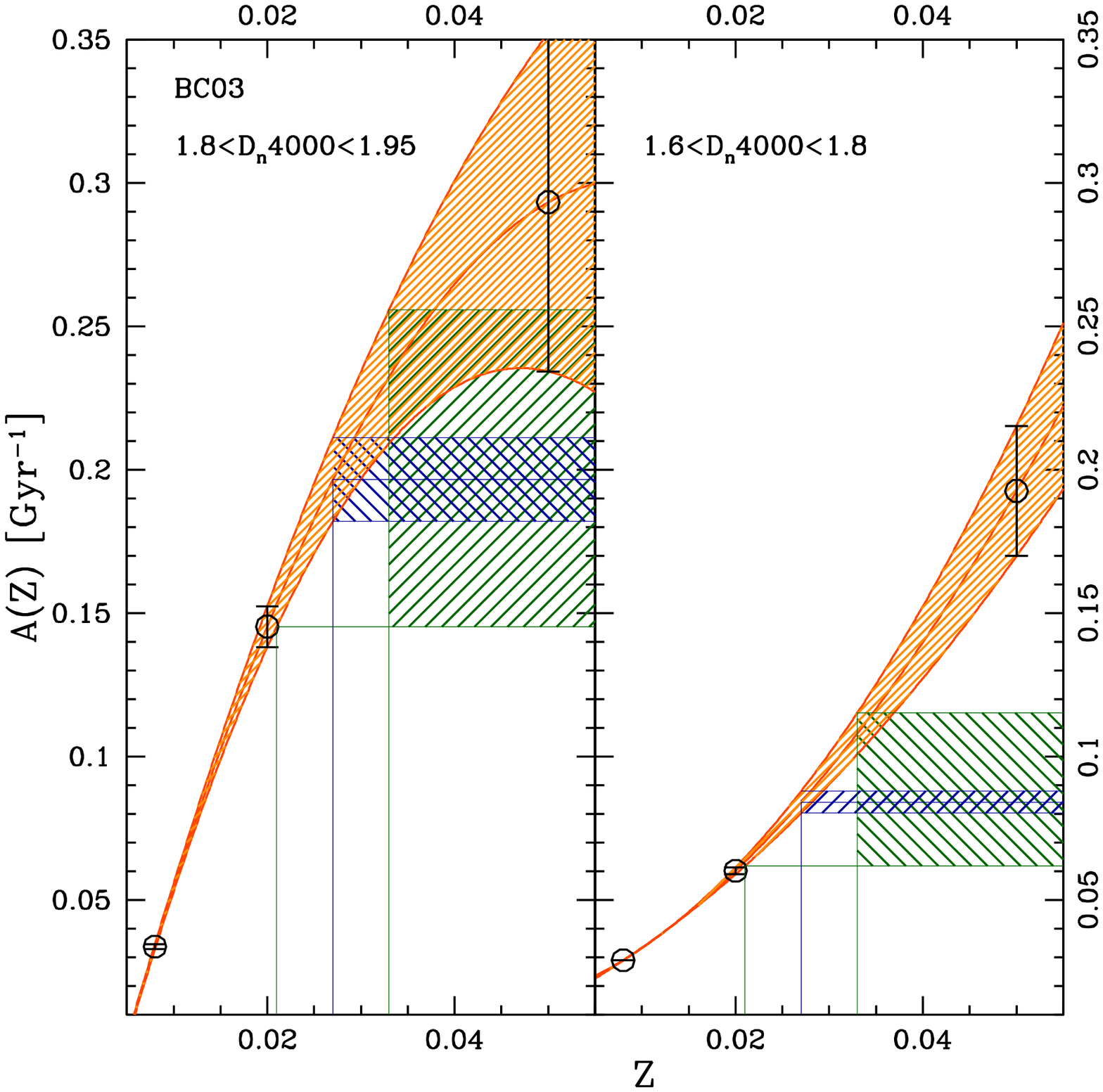}
\caption{Calibration of the $D_{n}4000$-age relations. Upper plots: $D_{n}4000$-age relation for different EPS models, metallicities and SFH. Left and right panels show M11 and
BC03 models, respectively, where the green lines show the relations for solar metallicity, blue for sub-solar and orange for over-solar;
lines from solid to dashed represent SFH with progressively higher $\tau$, from 0.05 to 0.3 Gyr. Lower panel show the linear best fit to
each metallicity, in the two $D_{n}4000$ ranges discussed in the text. Lower plots: fitted $A(SFH,Z/Z_{\odot})$-metallicity relation
for different EPS models. The orange shaded area represent the best fit to the data, shown as black points. The green shaded area 
represent the total error on $A(SFH,Z/Z_{\odot})$, given the uncertainty on the measured metallicity, while the blue area show the contribution to the error
due to SFH.
\label{fig:D4000mod}}
\end{center}
\end{figure}

Sigma-clipping has been applied to both red and blue $D_{n}4000$ bands of the spectrum, to remove contamination of 
residual night sky emission lines (see figure \ref{fig:sky}). This method provided results compatible with other approaches, but is more generic and less 
dependent on various assumptions (see appendix \ref{sec:appsky}). The BOSS spectrograph is composed of two separate instruments, a red and a a blue arm, connected at 
$\lambda\sim6000$~\AA. Any mis-calibration would impact the $D_{n}4000$ feature at $z\sim0.5-0.6$. Indeed a strong 
contamination (not clearly evident by visual inspection, but well detectable in the median relations) persists in all 
subsamples in the form of a wiggle of the $D_{n}4000$--z relations at $0.5<z<0.65$. For this reason we decided to 
restrict our analysis to $z<0.5$ (see figure \ref{fig:throughput} in the Appendix). A more detailed discussion can be found in appendix \ref{sec:appspectro}.

The median $D_{n}4000$--z relations are shown in the upper panel of figure \ref{fig:HzD4000}. These relations show a clear pattern, where more massive
galaxies present a larger break, confirming what was found in the analysis of ref.~\cite{Moresco2011}. This result can be interpreted in terms of redshift of 
formation, in particular for the three bins with the higher velocity dispersions where a smaller difference in metallicity is noticeable. For these samples it can 
be inferred that more massive galaxies are older and have formed their stellar mass at higher redshifts with respect to less massive ones, providing another 
observational confirmation to the mass-downsizing scenario \cite{Fontana2004,Thomas2005,Cimatti2007,Cimatti2009,Thomas2010}.

We estimated the quantity $A(SFH, Z/Z_{\odot})$ of eq. \ref{eq:HzD4000} following the approach discussed in ref.~\cite{Moresco2012a}. In that work, it was demonstrated that dividing the $D_{n}4000$ range into two regimes 
($1.6<D_{n}4000<1.8$ and $1.8<D_{n}4000<1.95$), the approximation of a linear relation between $D_{n}4000$ and age is extremely accurate independently of the considered model.
In particular, for each EPS model three metallicities (sub-solar, solar and over-solar) and four different SFHs (characteristic of the selected passive
population, i.e. a delayed exponentially declining SFH has been chosen, with $\tau=0.05, 0.1, 0.2, 0.3$ Gyr) are considered, and, at fixed metallicity, the slope of the $D_{n}4000$--age relation is obtained by averaging the values obtained for the various SFHs. The dispersion between the different SFHs has been taken into account as the associated error to 
$A(SFH,Z/Z_{\odot})$; then, the obtained values are interpolated to obtain the correct $A(SFH,Z/Z_{\odot})$ parameter at the metallicity of the sample.

The results for M11 and BC03 models are listed in Tab.~\ref{tab:Aparam} and shown in 
Fig. \ref{fig:D4000mod}. The upper panels show the $D_{n}4000$--age relations at different metallicities and SFHs for both M11 and BC03 models.
As it is evident from the bottom panels of the upper plots, in both $D_{n}4000$ ranges the linear fit to the data represent a very good approximation, with Spearman correlation coefficients always above 0.97, and on average of $\langle r\rangle=0.995\pm0.002$ in the lower $D_{n}4000$ range 
and $\langle r\rangle=0.9985\pm0.0003$ in the higher $D_{n}4000$ range for M11, and of $\langle r\rangle=0.989\pm0.003$
and $\langle r\rangle=0.9987\pm0.0004$ for BC03, respectively in the lower and higher $D_{n}4000$ range.
The lower panels of Fig. \ref{fig:D4000mod} show the factor $A(SFH,Z/Z_{\odot})$ as a function of the metallicity,
and the interpolation performed. Green and blue shaded area represent the total uncertainty on $A(SFH,Z/Z_{\odot})$ 
(given the metallicity range as measured from our sample, see Sect. \ref{sec:metallicity}), and the contribution due to SFH alone.
It is clear that the dominant systematic is at the moment due to our uncertainty on stellar metallicity, which contributes to
$\sim$80\% of the systematic error budget in M11, and to $\sim$75-85\% in BC03, depending on the $D_{n}4000$ range.

\begin{table}[t!]
\begin{center}
\begin{tabular}{|llcc|}
\hline \hline
& & $1.6<D_{n}4000<1.8$ & $1.8<D_{n}4000<1.95$\\
\hline
M11 & $A(Z/Z_{\odot}=0.5)$&$0.0299\pm0.0002$&$0.0305\pm0.0005$\\
M11 & $A(Z/Z_{\odot}=1)$&$0.065\pm 0.001$&$0.119\pm0.003$\\
M11 & $A(Z/Z_{\odot}=2)$&$0.138\pm0.02$&$0.22\pm0.02$\\
\hline
BC03 & $A(Z/Z_{\odot}=0.4)$&$0.02893\pm0.00004$&$0.037\pm0.001$\\
BC03 & $A(Z/Z_{\odot}=1)$&$0.0602\pm 0.001$&$0.145\pm0.007$\\
BC03 & $A(Z/Z_{\odot}=2.5)$&$0.193\pm0.002$&$0.29\pm0.06$\\
\hline \hline
\end{tabular}
\caption{$A(SFH,Z/Z_{\odot}))$ obtained from eq.~\ref{eq:D4000age} calibrating the $D_{n}4000$-age relation (in units of [Gyr$^{-1}$] 
for the M11 and BC03 models with different metallicities and for different $D_{n}4000$ ranges. The quoted errors are 
the dispersion between the values evaluated for different SFH choices, as discussed in the text.}
\label{tab:Aparam}
\end{center}
\end{table}

\begin{figure}[t!]
\begin{center}
\includegraphics[angle=-90, width=0.95\textwidth]{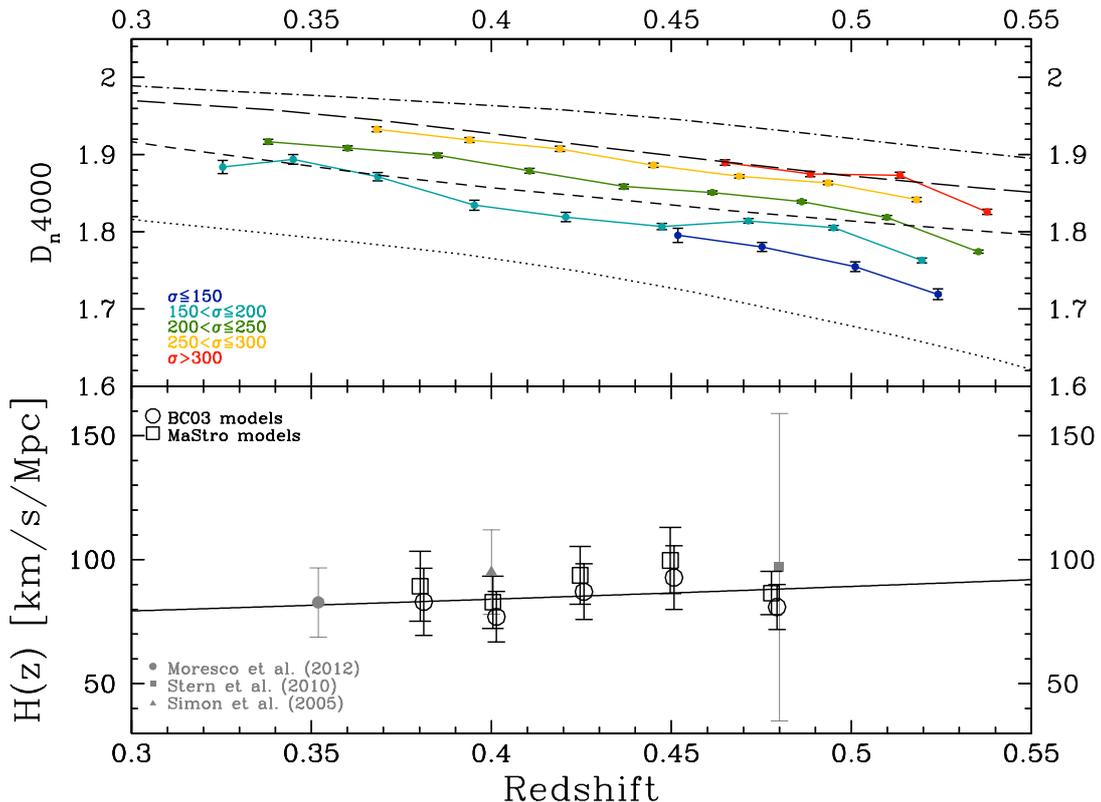}
\caption{Upper panel: median $D_{n}4000$--z relations obtained for the various velocity dispersion subsamples. The dashed lines show the theoretical $D_{n}4000$--z
relations estimated from M11 models (with solar metallicity) for four differed redshifts of formation, respectively 1, 1.5, 2, 2.5 from bottom to top. Lower panel: $H(z)$ 
measurements obtained with BC03 and M11 models, compared with literature data available in this redshift range \cite{Simon2005,Stern2010,Moresco2012a}. 
For illustrative purpose, the estimates obtained with BC03 models have been slightly offset in redshift.
\label{fig:HzD4000}}
\end{center}
\end{figure}

\section{Results}
\label{sec:results}
\subsection{The estimate of $H(z)$}
\label{sec:Hz}
We estimated the Hubble parameter $H(z)$ from eq. \ref{eq:HzD4000}. 
To keep under control the systematic uncertainties, we performed the analysis separately with two different EPS
models, BC03 and M11, and then compared the results.
The method to estimate $A(SFH,Z/Z_{\odot})$ has been discussed 
in section \ref{sec:D4000meas}. The stellar metallicity has been taken from the analysis presented in section \ref{sec:metallicity}, $Z/Z_{\odot}=1.35\pm0.3$. 
The differential $dz/dD_{n}4000$ has been obtained from the median $D_{n}4000$--z relation presented in the upper 
panel of figure \ref{fig:HzD4000}, considering the difference between the $i$-th and the $(i+N)$-th point for each velocity dispersion subsample. 
We have chosen $N=4$ as a trade-off to minimize the statistical scatter of the data over the intrinsic $D_{n}4000$ evolution with redshift, simultaneously 
maximizing the redshift sampling of $H(z)$. We therefore consider a redshift interval $\Delta z=0.1$, which corresponds to a difference in cosmic time 
$\sim0.7$ Gyr at $z\sim0.45$. We checked that different choices of $N$ (provided not to be dominated by the statistical scatter of the data) do not affect the results.

\begin{table}[t!]
\begin{center}
\begin{tabular}{||c||ccc|cc||ccc|cc||}
\hline
& \multicolumn{5}{|c||}{M11 models} & \multicolumn{5}{c||}{BC03 models} \\
$z$ & $H(z)$ & $\sigma_{stat}$ & $\sigma_{syst}$ & $\sigma_{tot}$ & \% error & $H(z)$ & $\sigma_{stat}$ & $\sigma_{syst}$ & $\sigma_{tot}$ & \% error\\
\hline
0.3802 & 89.3 & 3.2 & 13.7 & 14.1 & 15.8\% & 83.0 & 4.3 & 12.9 & 13.5 & 16.3\% \\
0.4004 & 82.8 & 2.4 & 10.3 & 10.6 & 12.8\% & 77.0 & 2.1 & 10 & 10.2 & 13.2\% \\
0.4247 & 93.7 & 2.7 & 11.4 & 11.7 & 12.4\% & 87.1 & 2.4 & 11 & 11.2 & 12.9\% \\
0.4497 & 99.7 & 3.1 & 13 & 13.4 & 13.4\% & 92.8 & 4.5 & 12.1 & 12.9 & 13.9\% \\
0.4783 & 86.6 & 2 & 8.5 & 8.7 & 10.1\% & 80.9 & 2.1 & 8.8 & 9 & 11.2\% \\
\hline \hline
$\langle 0.4293\rangle$ & 91.8 & 1 & 5.1 & 5.3 & 5.8\% & 85.7 & 1 & 5.1 & 5.2 & 6.1\% \\
\hline
\end{tabular}
\caption{$H(z)$ measurements (in units of [km/Mpc/s]) and their errors. The relative contribution of statistical and systematic errors
are reported, as well as the total error (estimated by summing in quadrature 
$\sigma_{stat}$ and $\sigma_{syst}$). These values have been estimated with M11 and BC03 EPS models respectively. 
For each model the averaged measurement is also reported. This dataset can be downloaded from 
http://www.physics-astronomy.unibo.it/en/research/areas/astrophysics/cosmology-with-cosmic-chronometers.}
\label{tab:Hz}
\end{center}
\end{table}

\begin{figure}[t!]
\begin{center}
\includegraphics[angle=-90, width=0.95\textwidth]{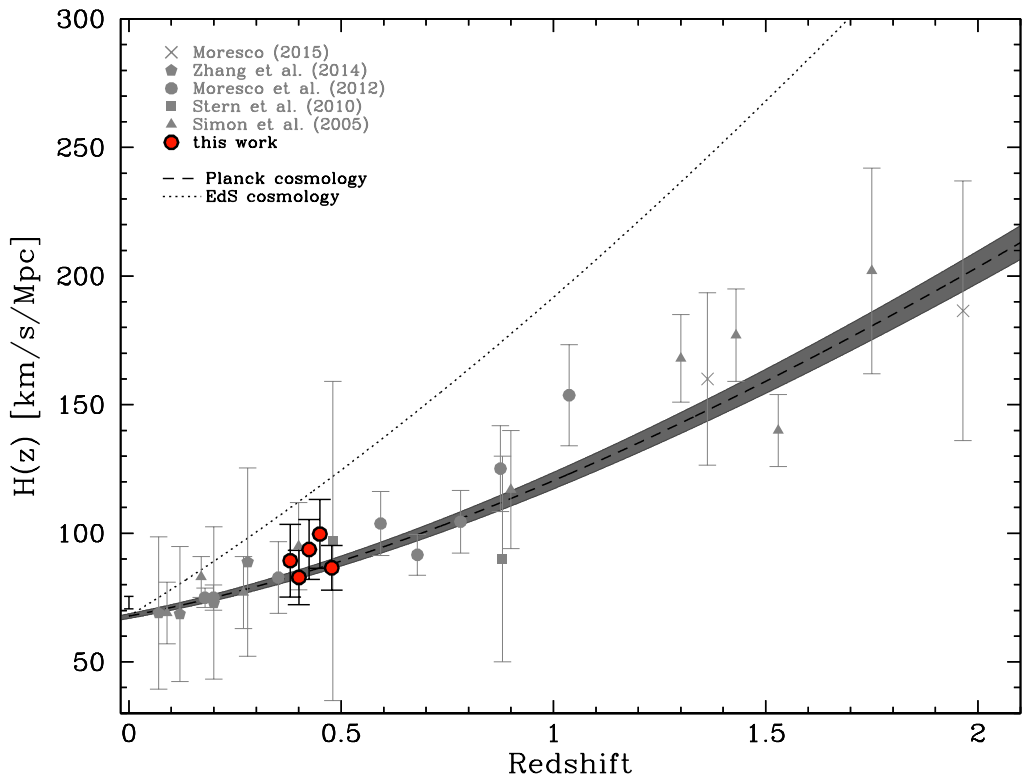}
\caption{Hubble parameter constraints obtained with M11 models, compared with various literature data \cite{Simon2005,Stern2010,Moresco2012a,Zhang2014,Moresco2015}.
The black point at $z=0$ is the Hubble constant constraints from ref.~\cite{Riess2011} with the recalibration of the distance to NGC 4258 from ref.~\cite{Humphreys2013}, i.e 
$H_{0}=73\pm2.4$ km/s/Mpc. The solid line and the dashed regions are not a fit to the data, but show the fiducial flat $\Lambda$CDM cosmology with its 1$\sigma$ uncertainty 
as constrained by Planck collaboration (\cite{Planck2015}, $H_{0}=67.8\pm0.9$ km/s/Mpc, $\Omega_{m}=0.308\pm0.012$). For comparison an Einstein-de Sitter 
model is shown, normalized to the same $H_{0}$.
\label{fig:Hz2}}
\end{center}
\end{figure}

The $H(z)$ measurements extracted from the different $\sigma$ subsamples are in good agreement, and, being statistically 
independent, they have been combined together averaging them in bins of $\Delta z=0.025$; we use a weighted 
mean where the weights are the corresponding error of each measurement. The results are shown in figure \ref{fig:HzD4000}, 
and presented in Tab.~\ref{tab:Hz}. Exploiting BOSS data, we are able to map homogeneously for the first time
the redshift range $0.3<z<0.5$ with an accuracy between 11\% and 16\%, which, as can be seen by figure \ref{fig:HzD4000}, 
was previously only poorly covered. The comparison between the results obtained with the two different EPS models also shows a good 
agreement, always $<1\sigma$, confirming the robustness of the estimate as found in ref.~\cite{Moresco2012a}. Despite the 
larger statistic of this analysis with respect to the one of ref.~\cite{Moresco2012a}, however, we have similar errorbars. This is 
due to the fact that in this work we have larger uncertainty on the slope $dz/dD_{n}4000$ since we are using smaller redshift 
arm ($\Delta z=0.1$ instead of 0.3).
Precise measurements in this redshift range are crucial to better constrain the time at which our Universe passed from decelerated to accelerated expansion. 

Analyzing Tab.~\ref{tab:Hz}, it is clear that, with the exceptional statistics provided by the BOSS data, systematics are now
dominating over statistical errors. Since the systematics themselves are mainly dominated by the uncertainty on metallicity 
and by the EPS model, additional work is needed to improve on the current analysis.
We plan to address this issue by exploiting the present dataset in a following paper, with a full and detailed spectral analysis.

To exploit the full constraining power of BOSS data, we decide also to provide a single $H(z)$ measurement, in which all mass subsamples have been averaged as discussed above,
but in a single point. This measurement is clearly not independent of the results reported in Tab. \ref{tab:Hz}, and should not be used in combination with them. 
The resulting constraint adopting M11 models is $H(z=0.4293)=91.8\pm5.3$ km/s/Mpc, and $H(z=0.4293)=85.7\pm5.2$ km/s/Mpc using BC03 models, 
reaching a 6\% accuracy including both statistic and systematic errors. 
As a comparison, we report the results obtained from the BAO analysis in the final BOSS DR12 \cite{Cuesta2015}, where it is found 
$H(z=0.57)r_{\rm d}/r^{\rm fid}_{\rm d}=100.3\pm3.7$ km/s/Mpc and $H(z=0.32)r_{\rm d}/r^{\rm fid}_{\rm d}=79.2\pm5.6$ km/s/Mpc. Considering that
$r^{\rm fid}_{\rm d}=147.10$ Mpc and $r_{\rm d}=147.41\pm0.30$ from Planck analysis \cite{Planck2015}, these measurements correspond to $H(z=0.57)=100.1\pm 3.7$ and 
$H(z=0.32)=79.0\pm5.6$km/s/Mpc, respectively a 3.7\% and 7.1\% measurement.

\subsection{The measurement of the transition redshift}

The cosmological transition redshift $z_{t}$ is defined as the redshift which separates the accelerated and decelerated expansion phases 
of the Universe. The BOSS data map the redshift range close to this epoch of cosmic re-acceleration homogeneously for 
the first time, and therefore allow us to measure it with cosmic chronometers with high accuracy.
We constrained $z_{t}$ by fitting all the ``cosmic chronometers'' data available so far in the literature 
\cite{Simon2005,Stern2010,Moresco2012a,Zhang2014,Moresco2015}, including the ones obtained in this work, with a standard 
$\chi^{2}$ approach. In this analysis, we have considered the measurements obtained assuming BC03 models, for consistency 
with literature derivations, as presented in Tab.~\ref{tab:Hzall} and figure \ref{fig:Hz2}.

The transition redshift determination can be done in the standard (model-dependent) way, assuming a Friedmann-Robertson-Walker 
metric and general relativity (i.e. $\Lambda$CDM-like cosmological model), and in a model-independent way. 

First of all, we consider an open $\Lambda$CDM cosmology:
\begin{equation}
H(z)=H_{0}\left[\Omega_{m}(1+z)^{3}+\Omega_{k}(1+z)^{2}+\Omega_{\Lambda}\right]^{1/2} ,
\label{eq:Hztheor}
\end{equation}
where $\Omega_{k}=1-\Omega_{m}-\Omega_{\Lambda}$. The transition redshift can be written as:
\begin{equation}
z_{t}=\left[\frac{2\Omega_{\Lambda}}{\Omega_{m}}\right]^{1/3}-1.
\label{eq:ztr}
\end{equation}
Combining equations \ref{eq:Hztheor} and \ref{eq:ztr}, we obtain a relation $H(z)_{mod}=f(H_{0},\Omega_{m},z_{t}$).
We assumed a gaussian prior on the Hubble constant $H_{0}=73\pm2.4$ km/s/Mpc \cite{Riess2011,Humphreys2013,Cuesta2015}, which is a cosmology-independent, 
direct measurement. 
For the other parameters, we considered uniform priors $\Omega_{m}\in[0,1]$ and $z_{t}\in[0,1.5]$. 
The results are presented in figure \ref{fig:chi2}: in the left panel the one-dimensional marginalized $\chi^{2}$ for $z_{t}$ is shown, while the right 
panel shows the two-dimensional constraints in the $\Omega_{m}-z_{t}$ plane. 

We obtained $ z_{t}=0.64^{+0.11}_{-0.07}$, which is in good agreement with Planck (2015) results \cite{Planck2015} assuming an open $\Lambda$CDM 
cosmology\footnote{see http://wiki.cosmos.esa.int/planckpla2015/index.php/Cosmological\_Parameters.}, as shown in figure \ref{fig:chi2}.
Our measurement is slightly larger than the estimate from ref.~\cite{Riess2007} ($ z_{t}=0.43^{+0.27}_{-0.08}$ at 95\% confidence level) obtained using SNe, and more consistent 
with the larger value of ref.~\cite{Lima2012} ($ z_{t}=0.73^{+0.45}_{-0.13}$ at 95\% confidence level), and with the estimate provided by the analysis of BAO in the Ly$\alpha$ forest 
of BOSS quasars ($ z_{t}\sim0.7$, \cite{Busca2013}). This result is in good agreement with the estimate from an independent fit to SNe of ref.~\cite{Capozziello2015}, providing
$z_{t}=0.643^{+0.034}_{-0.030}$ with a class of models deviating from GR (assuming $\Omega_{m}=0.315$). Ref.~\cite{Capozziello2014} analyzed the constraints 
on another class of $f(R)$ gravity models, and by combining SNe, BAO and older $H(z)$ measurements they obtained $z_{t}=0.7679^{+0.1831}_{-0.1829}$.
Refs.~\cite{Farooq2013a,Farooq2013b}, analyzing older $H(z)$ measurements in combination with BAO measurements in different cosmological scenarios, found a mean value
of $z_{t}=0.74\pm0.05$. Recently, also ref.~\cite{Rani2015} analysed the age of galaxies, strong lensing and SNe obtaining a constraint on $z_{t}<1$, considering both a parametric and a non-parametric approach.
Note that previous results were obtained from the analysis of SNe, comprising SNLS and Union2 samples, BAO and from the latest CMB 
measurements from Planck satellite, while in our case the measurement comes only from cosmic chronometers data and the local $H(z)$ measurement.

\begin{table}[t!]
\begin{center}
\begin{tabular}{lllrr|}
\multicolumn{4}{c}{{}}\\
\hline \hline
$z$ & $H(z)$ & $\sigma_{H(z)}$ & ref.\\
\hline
0.07 & 69.0 & 19.6 & \cite{Zhang2014}\\
0.09 & 69 & 12 & \cite{Simon2005}\\
0.12 & 68.6 & 26.2 & \cite{Zhang2014}\\
0.17 & 83 & 8 & \cite{Simon2005}\\
0.179 & 75 & 4 & \cite{Moresco2012a}\\
0.199 & 75 & 5 & \cite{Moresco2012a}\\
0.20 & 72.9 & 29.6 & \cite{Zhang2014}\\
0.27 & 77 & 14 & \cite{Simon2005}\\
0.28 & 88.8 & 36.6 & \cite{Zhang2014}\\
0.352 & 83 & 14 & \cite{Moresco2012a}\\
0.3802 & 83 & 13.5 & this work\\
0.4 & 95 & 17 & \cite{Simon2005}\\
0.4004 & 77 & 10.2 & this work\\
0.4247 & 87.1 & 11.2 & this work\\
0.44497 & 92.8 & 12.9 & this work\\
\hline \hline
\end{tabular}
\begin{tabular}{|lllrr|}
\multicolumn{4}{c}{{\it continues}}\\
\hline \hline
$z$ & $H(z)$ & $\sigma_{H(z)}$ & ref.\\
\hline
0.4783 & 80.9 & 9 & this work\\
0.48 & 97 & 62 & \cite{Stern2010}\\
0.593 & 104 & 13 & \cite{Moresco2012a}\\
0.68 & 92 & 8 & \cite{Moresco2012a}\\
0.781 & 105 & 12 & \cite{Moresco2012a}\\
0.875 & 125 & 17 & \cite{Moresco2012a}\\
0.88 & 90 & 40 & \cite{Stern2010}\\
0.9 &  117 &  23 & \cite{Simon2005}\\
1.037 & 154 & 20 & \cite{Moresco2012a}\\
1.3 & 168 & 17 & \cite{Simon2005}\\
1.363 & 160 & 33.6 & \cite{Moresco2015}\\
1.43 & 177 & 18 & \cite{Simon2005}\\
1.53 & 140 & 14 & \cite{Simon2005}\\
1.75 & 202 & 40 & \cite{Simon2005}\\
1.965 & 186.5 & 50.4 & \cite{Moresco2015}\\
\hline \hline
\end{tabular}
\caption{$H(z)$ measurements (in units [km/s/Mpc]) used for the cosmological analysis, and their errors.}
\label{tab:Hzall}
\end{center}
\end{table}

\begin{figure}[t!]
\begin{center}
\includegraphics[angle=0, width=0.46\textwidth]{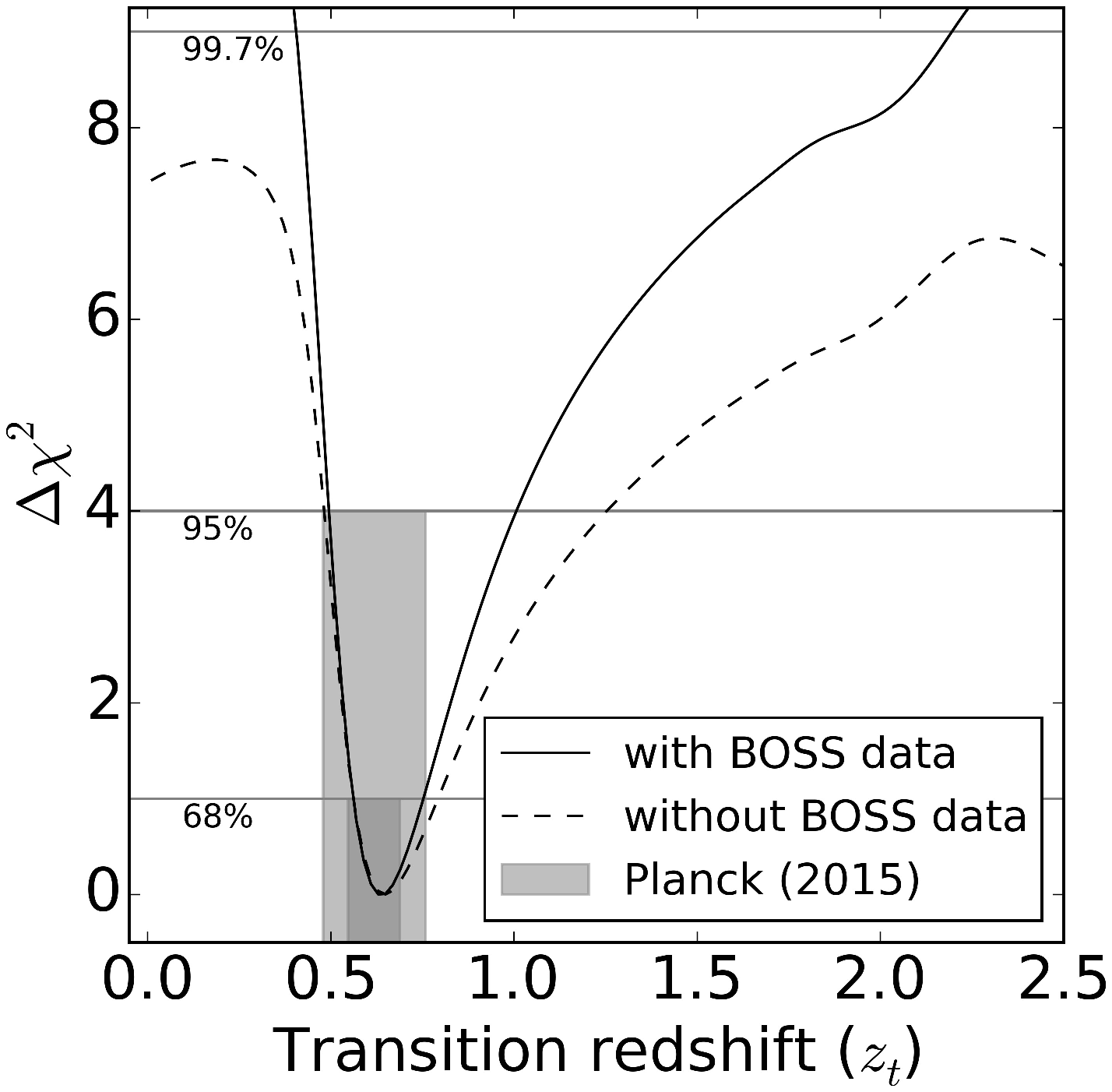}
\includegraphics[angle=0, width=0.47\textwidth]{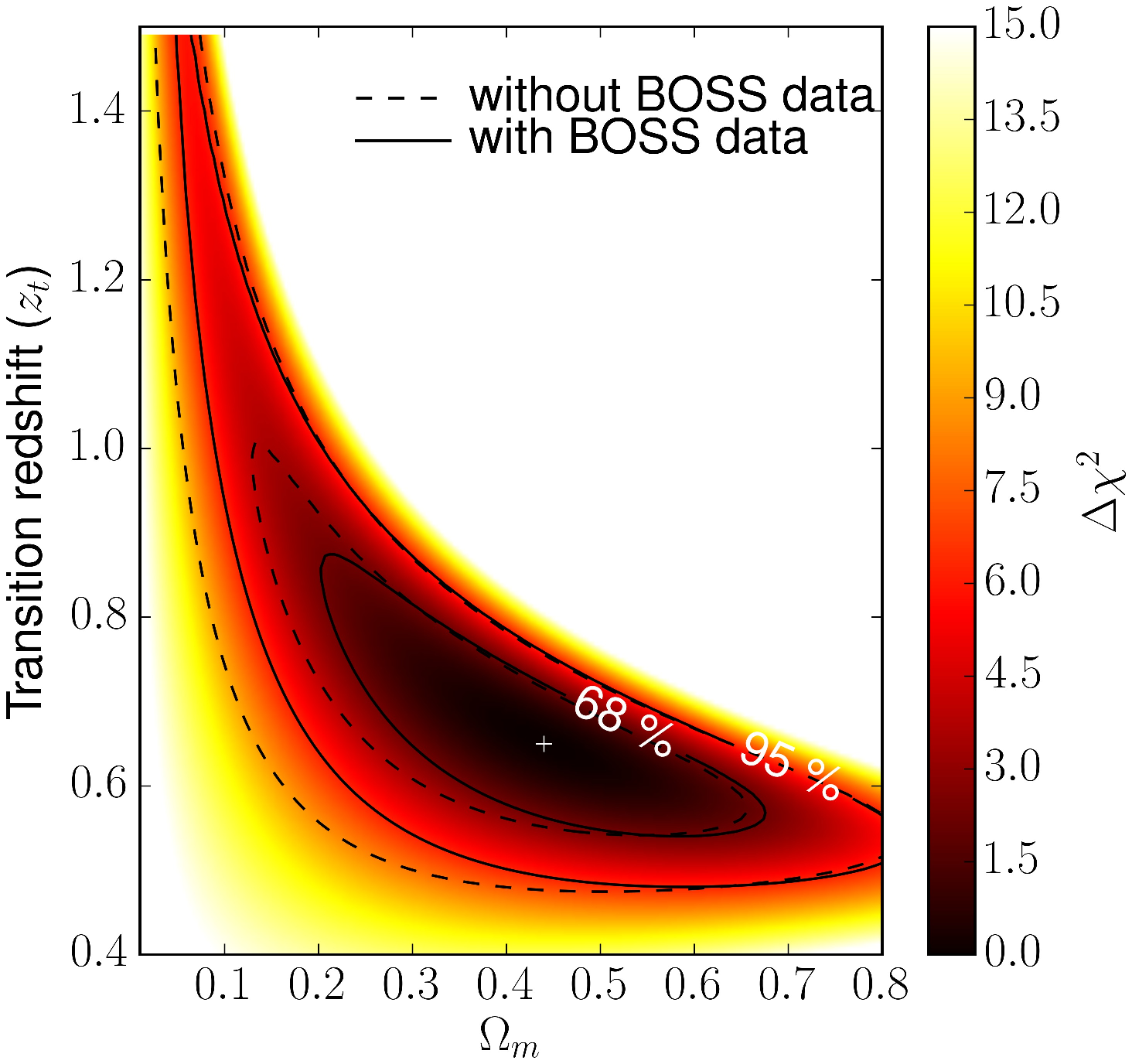}
\caption{Constraints on the cosmological transition redshift obtained in a o$\Lambda$CDM cosmology. The left panel shows the effective $\Delta\chi^{2}$
as a function of $z_{t}$. The solid line shows the constraint obtained with the data of Tab.~\ref{tab:Hzall}, while the dashed line shows the constraint obtained without 
the values obtained in this work. As a comparison, the grey shaded area represent the 1$\sigma$ and 2$\sigma$ constraints from Planck (2015) for an o$\Lambda$CDM cosmology. 
In the right panel the 68\% and 95\% two-dimensional constraints are shown in the $\Omega_{m}-z_{t}$ plane, both with and without the new $H(z)$ data obtained in this analysis.
\label{fig:chi2}}
\end{center}
\end{figure}

\begin{figure}[h!]
\begin{center}
\includegraphics[angle=-90, width=1.\textwidth]{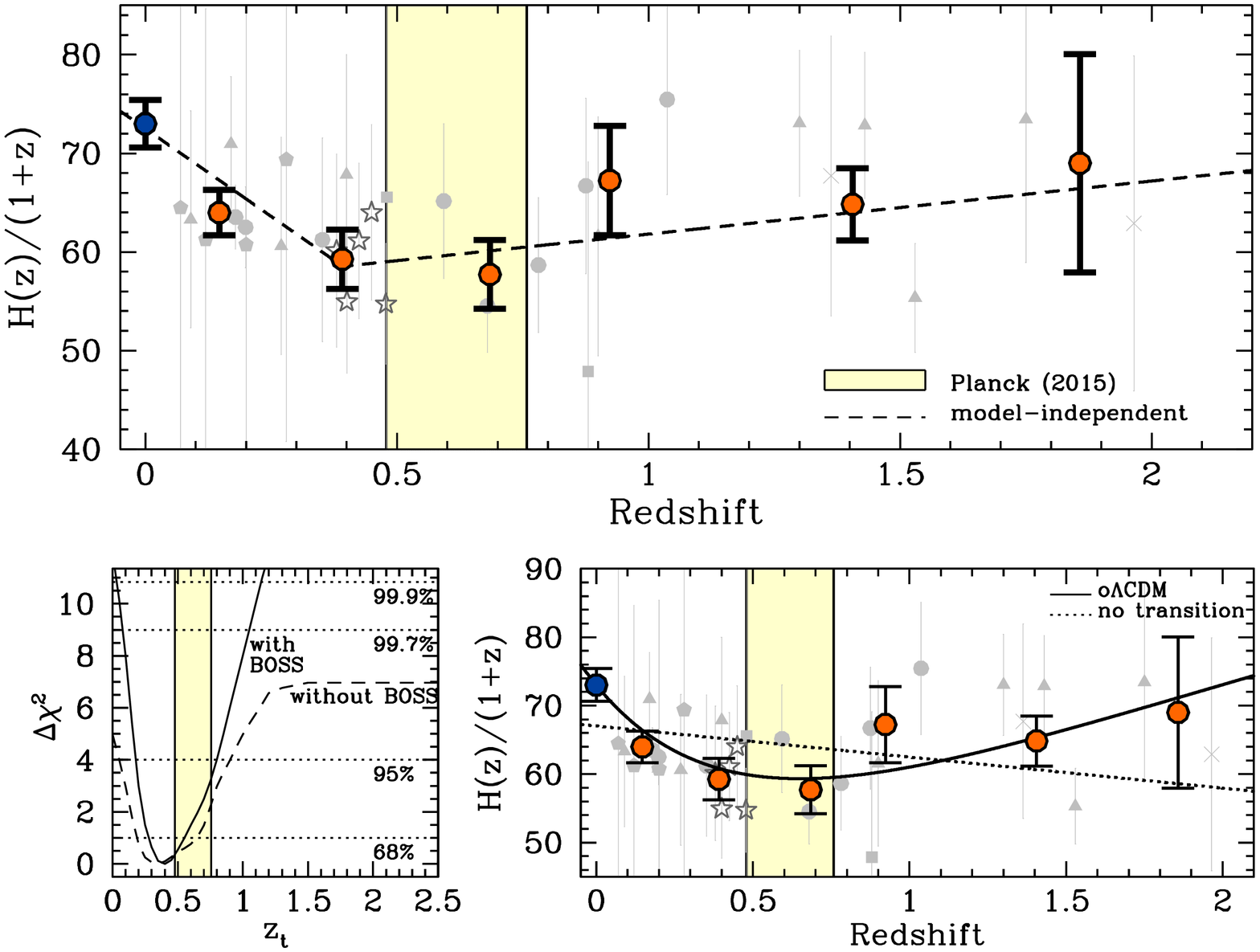}
\caption{Constraints on the cosmological transition redshift obtained in a model-independent way. In the upper panel are shown the fitted $H(z)/(1+z)$ measurements in grey, 
with the same coding of figure~\ref{fig:Hz2}, where the starred symbols are the measurements obtained from this analysis; for this fit we used BC03 
measurements for homogeneity with the estimates obtained in literature data. Orange points show the 
binned values, for illustrative purposes; the analysis is performed on unbinned data. The $H_{0}$ value used in this analysis is plotted in blue. 
The transition from decelerated to accelerated expansion is clearly visible by eye as a turn in the slope of the data. 
The solid line represents the best fit to the data with a cosmology-independent piecewise linear function. As a comparison, model dependent 2$\sigma$ 
constraints obtained from Planck (2015) \cite{Planck2015} in an open $\Lambda$CDM cosmology are shown in yellow.
In the lower-left panel the effective $\Delta\chi^{2}$ is shown as a function of $z_{t}$, with and without considering the new
BOSS data; the improvement of the statistical significance of the result provided by the new data is clearly evident. In the lower-right panel are presented 
for comparison the best fits obtained for the o$\Lambda$CDM model and for the model without a transition redshift.
\label{fig:chi2_mod}}
\end{center}
\end{figure}

The transition redshift can be also constrained without any further assumption from the function $f(z)=H(z)/(1+z)=\dot a$, that 
directly maps the acceleration of the Universe. This function has a positive slope when the Universe is accelerating, and a negative slope 
when it is decelerating. The deceleration parameter $q$ is related to $f(z)$ by $q = -H(z)/fÕ(z)$. 
Our measurements are crucial to provide the first cosmology-independent determination of the redshift of transition between decelerated and accelerated
expansion, $z_{t}$, at high statistical significance. Combining them with literature data as presented in Tab.~\ref{tab:Hzall} and adopting the most recent 
local distance ladder measurement of the Hubble constant $H_{0}$ \cite{Riess2011,Humphreys2013,Cuesta2015}, which is also independent of any 
cosmology-based constraint, we obtain 
31 $H(z)/(1+z)$ measurements (see figure \ref{fig:chi2_mod}). We fit these data with a piecewise linear function composed of two 
intervals (one for acceleration and one for deceleration), which is the simplest functional form parameterizing a change in the slope of $H(z)/(1+z)$ without 
assuming any model.
The transition redshift between the two regimes is a free parameter, together with the slope of the two lines and the overall amplitude. 
We use a standard $\chi^{2}$ approach, and, marginalizing over all other parameters, we find a transition redshift of $z_t = 0.4 \pm 0.1$ at the 68\% 
confidence level, as shown in figure ~\ref{fig:chi2_mod}.  We have also explored fitting with other functional forms like a parabola and a spline; 
however, the piecewise function provides the best $\chi^2$ and the transition redshift does not change within the 68\% confidence limit. 

We also attempt a fit to the data with a straight-line, where no transition between decelerated and accelerated expansion is assumed, and compare the results.
The BOSS data have proven to be crucial for the detection of the transition redshift in both model-dependent and model-independent measurements, as it is evident
from figure~\ref{fig:chi2} and \ref{fig:chi2_mod}. For the piecewise fit, without the new BOSS data we obtain no constraint on the transition redshift at a significance of 2.6$\sigma$ 
or higher within the redshift range probed by our sample, while on the contrary, considering the new data we find that the transition redshift  for the piecewise fit is 
$0.2<z_t<0.8$ at $2\sigma$ (95\%) and $0.1<z_t<1.05$ at 3$\sigma$ (99.7\%) levels, with a detection of $0<z_{t}<1.15$ at 99.9\% confidence level. Similarly, for the o$\Lambda$CDM fit 
we obtain no constraint at the 3$\sigma$ level (99.7\%) without the new data (with $z_t<3.5$), and $0.41<z_t<2.2$ considering them, proving that BOSS 
data are necessary for a significant detection.

Different estimators have been proposed in the literature to select the best model fitting a set of data; among them, the most common are the Akaike Information Criterion 
(AIC, \cite{Akaike1974}), and the Bayesian Information Criterion (BIC, \cite{Schwarz1978}). These criteria indicate which is the model preferred by data, 
and the level of confidence at which other models can be discarded, implementing an ``Occam razor'' selection that penalizes increasing the number of parameters. 
In this way, they penalize over-fitting the data (for more details, see \cite{Liddle2007}). 
Given $k$ degrees of freedom in a model, and N data points, they are defined as:
\begin{equation}
AIC=-2 \ln \mathcal{L}_{max} + 2k,\;\;\;\;\;\;\;\;BIC=-2 \ln \mathcal{L}_{max} + k \ln {\rm N}.
\end{equation}
The BIC penalizes an increased number of free parameters more than AIC; moreover, ref.~\cite{Sugiura1978} proposed a correction to AIC which is more correct for
finite number of data points:
\begin{equation}
AIC_{c}=AIC+\frac{2k(k+1)}{N-k-1}.
\end{equation}
Different models are then compared in terms of their $\Delta AIC_{c}$ or $\Delta BIC$ based on the Jeffreys' scale \cite{Jeffreys1961}, so that a difference $>5$
is highly significant. In tab.~\ref{tab:res_zt} are summarized the results. With both estimators, we find that the model preferred
by the data is the o$\Lambda$CDM, but without significant difference with respect to the cosmology-independent linear piecewise fit. On the contrary, 
a straight-line is strongly disfavored by the data, both considering $AIC_{c}$ and $BIC$. 
\begin{table}[t!]
\begin{center}
\begin{tabular}{ccccc}
\hline
\hline
model & assumption & result & AICc & BIC\\
\hline
o$\Lambda$CDM & cosmology & $ z_{t}=0.64^{+0.1}_{-0.06}$ & 21.41 & 25.71\\ 
piecewise linear fit & model-independent & $ z_{t}=0.4\pm0.1$ & 23.99 & 28.72\\
linear fit & model-independent & -- & 32.42 & 35.29\\
\hline
\hline
\end{tabular}
\caption{Constraints on the cosmological transition redshift $ z_{t}$ obtained by fitting cosmic chronometer data
with different models. The results are presented at 68\% confidence level; all constraints assume a gaussian prior on the Hubble constant 
$H_{0}=73\pm2.4$ km/s/Mpc \cite{Riess2011,Humphreys2013,Cuesta2015}.}
\label{tab:res_zt}
\end{center}
\end{table}
The model with a transition redshift is preferred at highly 
significant level compared to a model with no transition redshift, therefore ruling out the null hypothesis of no redshift of transition at {\color{black}99.9\% confidence level}.

We have shown that the choice of stellar population models does not affect the derived $H(z)$ value and that differences are always smaller than $1\sigma$. However, this small 
systematic shift might have a larger impact in computing the transition redshift, as it is, effectively, a derivative of $H(z)$. In our data-set only 15 $H(z)$ measurements are 
derived with M11 models. For this sub-set, following the same approach described in section \ref{sec:Hz}, we find $z_t = 0.75\pm0.15$ (we note, however, that at 95\% confidence 
level the constraints degrade considerably because of the reduced number of points, giving $0.1 < z_t < 2$). As a further test, we also explored the full BC03 dataset, but 
using instead M11 $H(z)$ measurements where available. This could be thought as an extreme example of systematic impact, as one should use just a given EPS 
model where the homogeneity of the EPS model ensures that these differences cancel out when estimating the derivative. In this case, we find $z_t = 0.75\pm0.15$ and 
$0.25 < z_t < 1.0$ at 95\% confidence. This is perfectly consistent with our BC03 determination, but with a larger errorbar at higher redshifts. When using Bayesian 
evidence to analyze the significance of excluding the hypothesis of no transition redshift with this latest data-set, which combines estimates from different EPS models, 
we find a substantial-strong evidence in agreement with our general analysis. Our conclusion is that even an inhomogeneous analysis of the data provides a significant 
evidence for a transition redshift between deceleration and acceleration.

\section{Conclusions}
In this paper, we analyzed BOSS data to set constraints on the expansion history of the Universe
through the ``cosmic chronometers'' approach. We implemented the technique suggested in refs.~\cite{Moresco2011,Moresco2012a},
in which it was proposed to constrain the Hubble parameter $H(z)$ from the differential evolution
of a spectral feature, the $D_{n}4000$, of very massive and passive galaxies.

The main conclusion of this analysis can be
summarized as follows:
\begin{itemize}
\item We implement a strict selection criterion to extract, amongst BOSS-DR9, the most 
massive and passive sample, the least biased by star-forming emission-lines outliers. In this way we 
select more than 130000 massive and passive galaxies, which are binned in narrow redshift bins 
to follow the evolution of this population with cosmic time, and in velocity dispersion bins to follow this evolution 
in different mass bins.
\item For all the galaxies of our sample, we measure the $D_{n}4000$. We also create median stacked spectra
in all the velocity dispersion and redshift bins, which are analyzed with a full spectral fitting technique to extract
information on the stellar metallicity of the samples.
We apply three different full spectral fitting codes using two EPS models (BC03 and M11), to explore the dependence 
of the constraints on both the software implemented and the model adopted. These codes yield comparable results, 
suggesting for our sample an average metallicity of $Z/Z_{\odot}=1.35\pm0.3$, in agreement with other independent estimates for 
this population of galaxies.
\item The inspection of the median $D_{n}4000$--z relation provides evidence supporting the mass-downsizing
scenario, with more massive galaxies having a larger break and, given the metallicity constraints obtained,
older ages with respect to less massive galaxies.
\item These measurements are used to obtain five new cosmology-independent $H(z)$ constraints in the redshift range $0.3<z<0.5$,
with an accuracy $\sim$11-16\%, taking into account both statistical and systematic error. These constraints are
obtained adopting two different EPS models, BC03 and M11, to study the dependence of our results on models, finding
no significant difference. These new constraints allow us to homogeneously map for the first time this range of cosmic times,
which are crucial to disentangle the epoch which separates the decelerated and accelerated phases of the expansion of
the Universe.
\item To exploit the constraining power of BOSS data, we combine the five measurements, obtaining a constraint 
$H(z=0.4293)=91.8\pm5.3$ km/s/Mpc using M11 models, and $H(z\sim0.43)=85.7\pm5.2$ km/s/Mpc using 
BC03 models, a 6\% measurement including both statistical and systematic errors. This result is a complementary and cosmology-independent result that is comparable to the result
obtained from the analysis of BAO from the final BOSS-DR12 sample \cite{Cuesta2015}.
\item We use the new data obtained, jointly with other cosmic chronometers literature data, to set constraints on the cosmological
transition redshift, considering an open $\Lambda$CDM cosmology. We obtain a value $\rm z_{t}=0.64^{+0.11}_{-0.06}$ in perfect agreement 
with the estimate from Planck (2015) \cite{Planck2015} and SNe \cite{Lima2012, Capozziello2014,Capozziello2015} analyses.
\item The new data from SDSS-III/BOSS allow us to implement also an independent fit to the function $H(z)/(1+z)$, 
which directly probes the acceleration of the Universe, using a functional form without any further cosmological assumption. 
In this way, we obtain the constraint $z_{t}=0.4\pm0.1$. At 99.9\% confidence level, this analysis yield a detection of the transition redshift as $0<z_{t}<1.15$,
ruling out a straight-line fit, which would imply no transition from deceleration to acceleration, at very high significance. This is 
the first time that the epoch of cosmic re-acceleration has been observed in a cosmological model-independent way.
\end{itemize}

This work has shown the potential of the {\em cosmic chronometers} approach to set cosmology-independent
constraints on the Hubble parameter $H(z)$. This technique can be considered as a complementary tool
with respect to more standard ones (e.g. BAO, SNe) to set constraints on cosmological models, and
to keep the systematic affecting each probe under control. While it has been demonstrated that in many cases
the constraining power of this method is comparable with the power of classical approaches,
we also note that at the present status the measurements are systematic-dominated. Looking towards the future, on one hand we foresee that 
an improvement in the measurement of metallicity of passive
galaxies will represent a decisive step to minimize systematics, and significantly reduce errorbars; on the other hand,
it will be crucial to improve the samples at $z>0.5$, in order to maximize the accuracy of the measurements over a 
larger redshift range. From this point of view, future missions like Euclid \cite{Laurejis2011}, WFIRST \cite{Spergel2013},
DESI \cite{DESI} and LSST \cite{LSST} will represent an ideal starting point at these high redshifts.

In future papers, we plan to explore the capability of these new data, in combination with Planck (2015) measurements,
to set constraints on parameters in different cosmological scenarios, focusing on the additional constraining power
provided by cosmic chronometers data, and to take advantage of the high-quality stacked spectra obtained in this analysis
to set evolutionary constraints on the properties of passive galaxies.
  
\acknowledgments{MM, LP and AC acknowledge financial contributions by grants ASI/INAF
I/023/12/0 and PRIN MIUR 2010-2011 ``The dark Universe and the cosmic
evolution of baryons: from current surveys to Euclid''. 

RJ and LV thank the Royal Society for financial support and the ICIC at Imperial College for 
hospitality while this work was being completed. LV is supported by the European Research 
Council under the European Community's Seventh Framework Programme FP7-IDEAS-Phys.LSS 
240117. Funding for this work was partially provided by the Spanish MINECO under projects 
AYA2014-58747-P and MDM-2014-0369 of ICCUB (Unidad de Excelencia ``Maria de Maeztu'')

Funding for SDSS-III has been provided by the Alfred P. Sloan Foundation, the Participating 
Institutions, the National Science Foundation, and the U.S. Department of Energy Office of 
Science. The SDSS-III web site is http://www.sdss3.org/.

SDSS-III is managed by the Astrophysical Research Consortium for the Participating Institutions 
of the SDSS-III Collaboration including the University of Arizona, the Brazilian Participation Group, 
Brookhaven National Laboratory, Carnegie Mellon University, University of Florida, the French 
Participation Group, the German Participation Group, Harvard University, the Instituto de Astrofisica 
de Canarias, the Michigan State/Notre Dame/JINA Participation Group, Johns Hopkins University, 
Lawrence Berkeley National Laboratory, Max Planck Institute for Astrophysics, Max Planck Institute 
for Extraterrestrial Physics, New Mexico State University, New York University, Ohio State University, 
Pennsylvania State University, University of Portsmouth, Princeton University, the Spanish Participation 
Group, University of Tokyo, University of Utah, Vanderbilt University, University of Virginia, University 
of Washington, and Yale University.}

\appendix

\section{Correcting for night sky emission lines residuals}
\label{sec:appsky}

\begin{figure}[b!]
\begin{center}
\includegraphics[angle=0, width=0.85\textwidth]{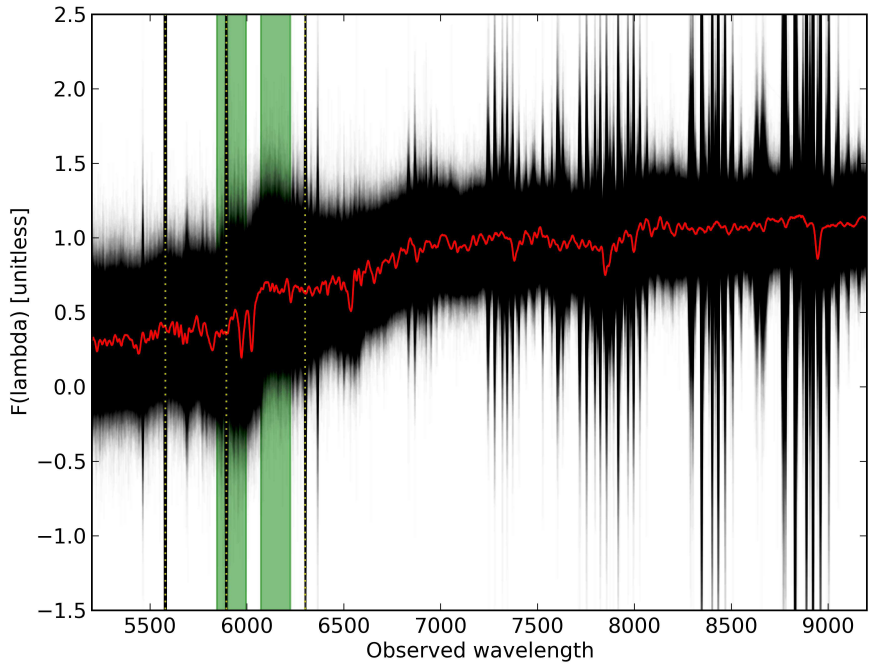}
\caption{Example of median stacked spectrum (in the observed frame) at $z=0.5181$ for the subsample with ${\rm 250<\sigma<300}$ km/s.
The median spectrum is shown in red, while in black are superimposed all the spectra used to create the stack. Particularly evident
are the residual of the 5577~\AA, of the 5892~\AA~and of the 6300~\AA~lines, highlighted with the yellow dotted lines. 
In green are shown the regions of the red and blue bands of the $D_{n}4000$.
\label{fig:sky}}
\end{center}
\end{figure}

Night sky line residuals can be important source of noise in (BOSS) spectra. As discussed here and within the BOSS collaboration\footnote{See 
https://www.sdss3.org/dr9/spectro/caveats.php\#night\_sky.}, the lines at 5577~\AA, 6300~\AA, and 6363~\AA~in particular may leave significant 
spikes in the spectra, and bias the measurement of the break if they happen to be in $D_{n}4000$ red or blue band. Together with the previous lines, we 
identify in sky spectra five additional lines with significant residuals at 5892~\AA, 5917~\AA, 5934~\AA~and 5955~\AA, as shown in figure 
\ref{fig:sky}. These residuals bias the measurement of $D_{n}4000$, leaving an imprint in the median $D_{n}4000$--z relation in the form of wiggles 
(particularly at $z\sim0.4$) that depart from the expected theoretical behavior, as can be seen in the left plot of figure \ref{fig:throughput}.
To overcome this issue, we applied a sigma-clipping procedure to the observed spectra in the red and blue bands of $D_{n}4000$,
removing pixels with a $>4\sigma$ difference from the mean. This helps in particular to remove the wiggle at $z\sim0.4$, where the strong
5577~\AA~line falls in the $D_{n}4000$ range.

We also explored different methods to clean the spectra from the residuals of the sky emission lines, namely cutting the contaminated pixels,
and weighting the pixels with their variance. Each method gave similar results to the one obtained with the sigma clipping approach, being however
more dependent on various assumptions (such as how many pixels to cut, how to treat the removed regions, and how to properly weight the pixels).
We therefore decide to adopt the first method, and applied it to the data.

\begin{figure}[t!]
\begin{center}
\includegraphics[angle=-90, width=0.49\textwidth]{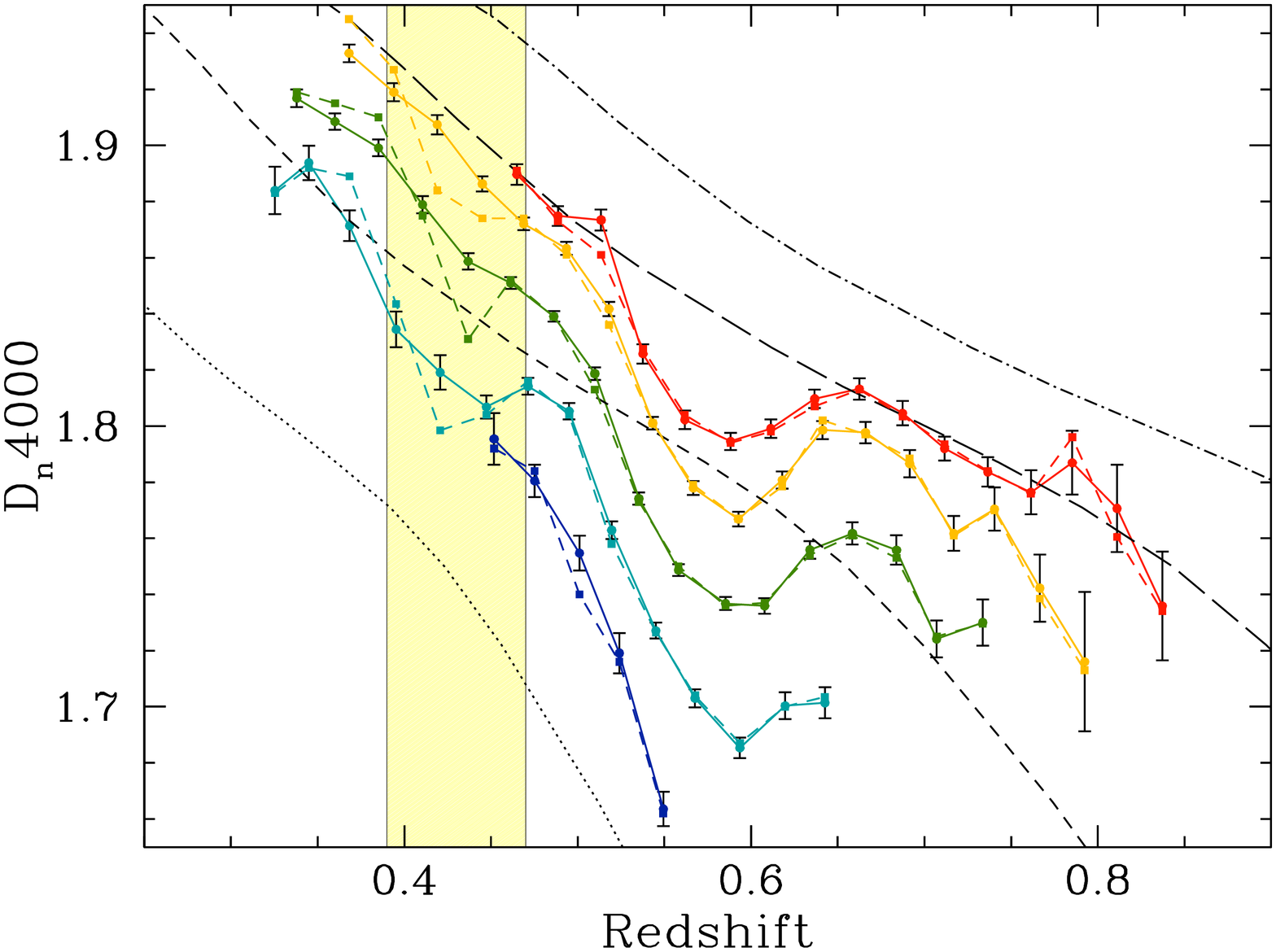}
\includegraphics[angle=-90, width=0.49\textwidth]{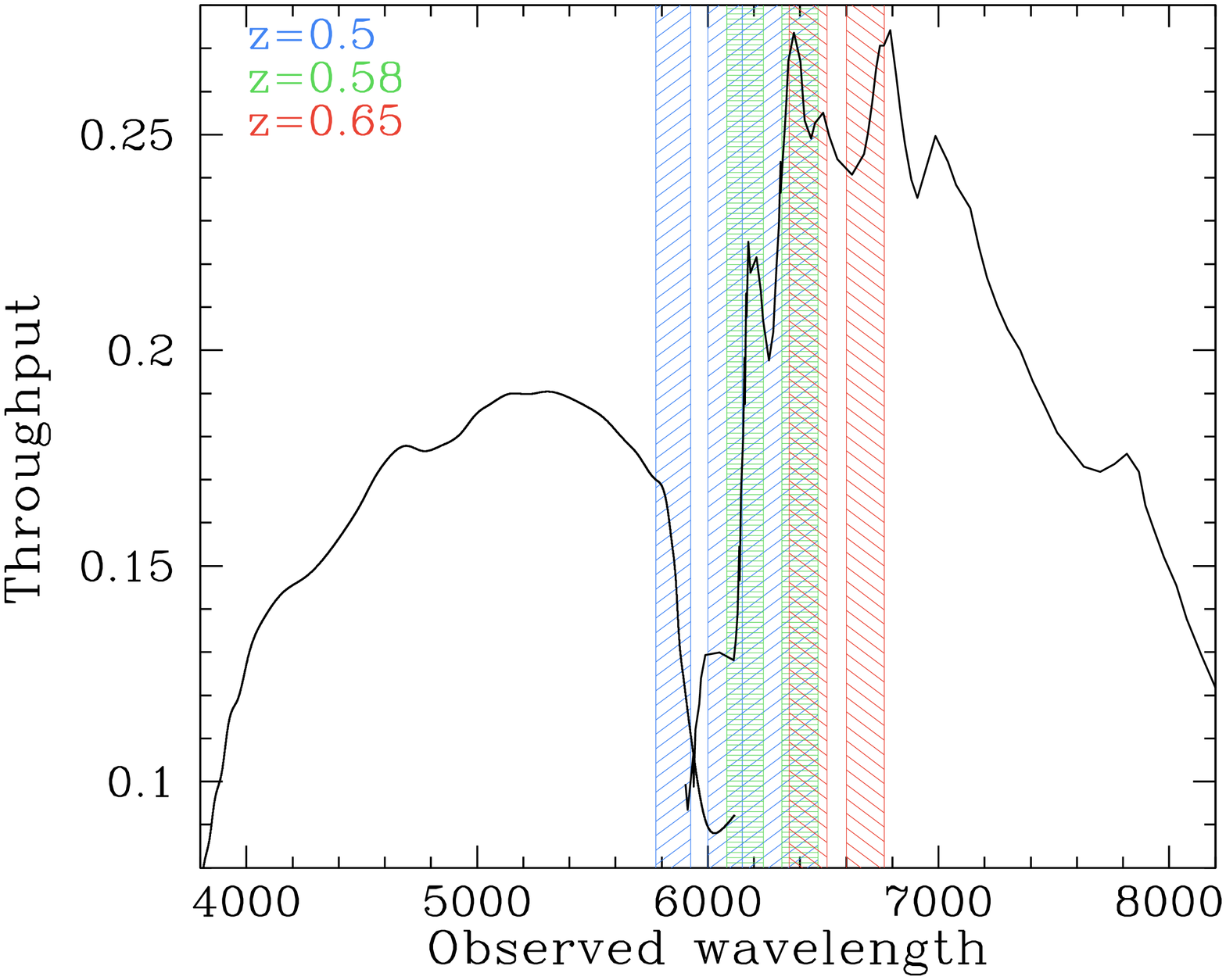}
\caption{In the left plot is shown the median $D_{n}4000$--z relation before (open points, dotted lines) and after (filled points, solid
lines) the sigma clipping correction; the correction is in particular effective around $z\sim0.4$, where the 5577~\AA~line enters the $D_{n}4000$ range. As
in figure \ref{fig:HzD4000}, dashed lines show $D_{n}4000$ from models.
In the right plot are presented the throughput curves for the red and blue arms of BOSS spectrograph (\cite{Smee2013}, see also 
http://www.sdss.org/dr12/algorithms/spectrophotometry/) as a function of observed wavelength. The shaded regions with the same colors highlight the left 
and right bands where $D_{n}4000$ is calculated, at three different redshifts ($z=0.5$ in red, $z=0.58$ in green, adn $z=0.65$ in blue) around which the wiggle 
in the median $D_{n}4000$--z relations is observed, showing that at these redshifts the $D_{n}4000$ transits from the blue to the red arm of the spectrograph.
\label{fig:throughput}}
\end{center}
\end{figure}

\section{BOSS throughput analysis}
\label{sec:appspectro}
The BOSS spectra are obtained with two separate instruments, a blue arm and a red arm, with an overlapping region around $\sim$6000~\AA, as shown in the 
right plot of figure \ref{fig:throughput}. This observed range falls exactly in the $D_{n}4000$ range, once redshifted at $z\sim0.5$, and in particular we notice a 
correlation between the descending and ascending trend of the wiggle (described in Appendix A) and the $D_{n}4000$ being measured on the blue arm, in the red arm, or in between 
(see right plot of figure \ref{fig:throughput}).
This correlation suggests a possible sub-optimal calibration between the two spectrographs, that results in the observed wiggle in the median $D_{n}4000$--z
relation. Even if this wiggle is actually small, of the order of $\Delta D_{n}4000\sim0.05$, it significantly affects the $H(z)$ measurement, that critically depends
on the differential $dD_{n}4000$. In order to avoid biasing our measurements, we therefore decided to restrict our analysis to $z<0.5$.

\bibliographystyle{ieeetr}
\bibliography{bib}

\end{document}